\documentclass{sig-alternate-05-2015}
\setcopyright{acmcopyright}
\usepackage[english]{babel}
\usepackage[latin1]{inputenc}
\usepackage{amssymb}
\usepackage[ruled]{algorithm2e}
\usepackage{algpseudocode}
\usepackage{bm}
\usepackage{amsmath}
\usepackage{graphicx}
\usepackage{epsfig}
\usepackage{url}
\usepackage{psfrag}
\usepackage{balance}
\usepackage{pstricks}
\usepackage{times}
\usepackage{array}
\usepackage{subfigure}
\usepackage{float}
\usepackage{multirow}
\usepackage{mathtools}
\usepackage[sort,nospace,noadjust]{cite}
\graphicspath{{.}{figs/}}

\acmPrice{\$15.00}

\newtheorem{theorem}{\textbf{Theorem}}

\newcommand{\tabincell}[2]{\begin{tabular}{@{}#1@{}}#2\end{tabular}}

\begin{document}

\title{A Fast Sampling Method of Exploring Graphlet Degrees of Large Directed and Undirected Graphs}
\numberofauthors{1}
\author{
\alignauthor
Pinghui Wang$^{1}$, Xiangliang Zhang$^{2}$, Zhenguo Li$^{3}$, Jiefeng Cheng$^{3}$,\\
John C.S. Lui$^{4}$, Don Towsley$^{5}$, Junzhou Zhao$^{4}$, Jing Tao$^{1}$, and Xiaohong Guan$^{1,6}$\\
{\small
\affaddr{$^{1}$MOE Key Laboratory for Intelligent Networks and Network Security, Xi'an Jiaotong University, China}\\
\affaddr{$^{2}$King Abdullah University of Science and Technology, Thuwal, SA}\\
\affaddr{$^{3}$Huawei Noah's Ark Lab, Hong Kong}\\
\affaddr{$^{4}$Department of Computer Science and Engineering, The Chinese University of Hong Kong, Hong Kong}\\
\affaddr{$^{5}$Department of Computer Science, University of Massachusetts Amherst, MA, USA}\\
\affaddr{$^{6}$Department of Automation and NLIST Lab, Tsinghua University, Beijing, China}\\
\{phwang, jtao, jzzhao, xhguan\}@sei.xjtu.edu.cn, xiangliang.zhang@kaust.edu.sa,\\
\{li.zhenguo, cheng.jiefeng\}@huawei.com, cslui@cse.cuhk.edu.hk, towsley@cs.umass.edu
}
}

\maketitle

\begin{abstract}
Exploring small connected and induced subgraph patterns (CIS patterns, or graphlets) has recently attracted considerable attention.
Despite recent efforts on computing the number of instances
a specific graphlet appears in a large graph (i.e., the total number of CISes isomorphic to the graphlet),
little attention has been paid to characterizing a node's graphlet degree, i.e., the number of CISes isomorphic to the graphlet that include the node,
which is an important metric for analyzing complex networks such as social and biological networks.
Similar to global graphlet counting, it is challenging to compute node graphlet degrees for a large graph due to the combinatorial nature of the problem.
Unfortunately, previous methods of computing global graphlet counts are not suited to solve this problem.
In this paper we propose sampling methods to estimate node graphlet degrees for undirected and directed graphs,
and analyze the error of our estimates.
To the best of our knowledge, we are the first to study this problem and give a fast scalable solution.
We conduct experiments on a variety of real-word datasets
that demonstrate that our methods accurately and efficiently estimate node graphlet degrees for graphs with millions of edges.
\end{abstract}

\section{Introduction} \label{sec:introduction}
\begin{figure*}[htb]
\center
\subfigure[undirected graphlets and their orbits. \label{fig:GDDgraphlets}]{
\includegraphics[width=0.298\textwidth]{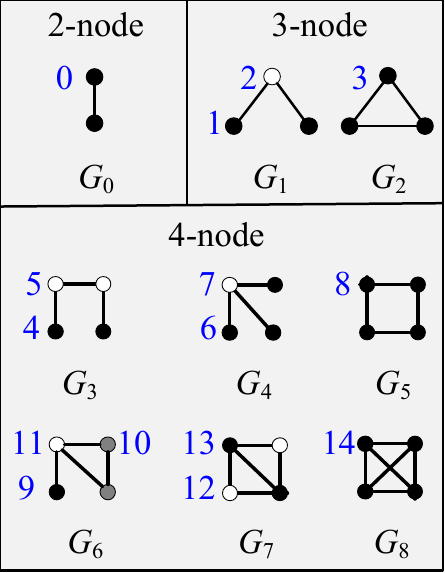}}
\subfigure[3-node directed graphlets and their orbits.\label{fig:othergraphlets}]{
\includegraphics[width=0.686\textwidth]{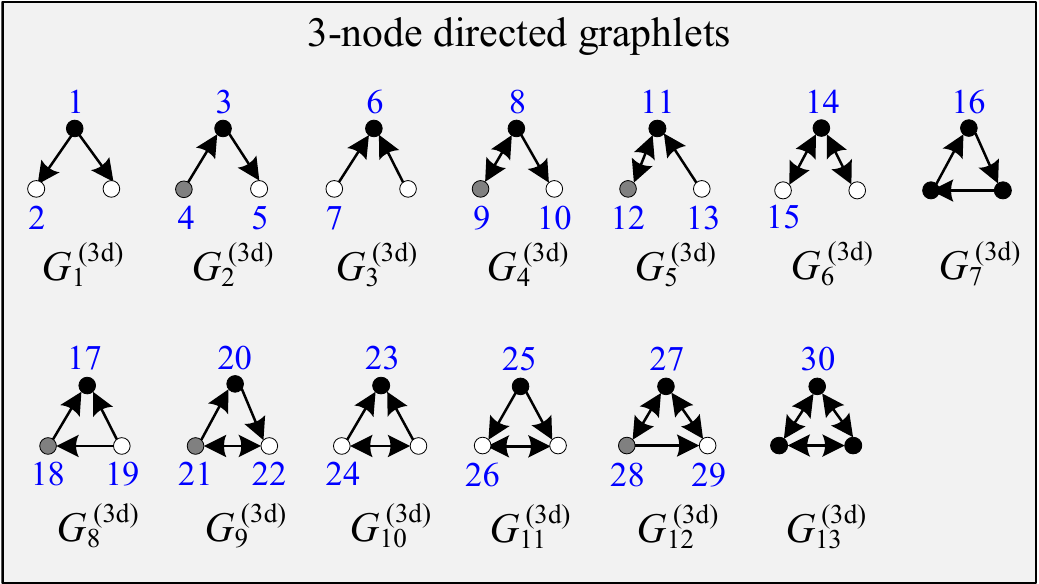}}
\caption{Graphlets and their automorphism orbits studied in this paper. Numbers in blue are orbit IDs.
There is one 2-node undirected graphlet $G_0$, two 3-node undirected graphlets $G_1$ and $G_2$, six 4-node undirected graphlets $G_3, \ldots G_8$,
and thirteen 3-node directed graphlets $G_1^{\text{(3d)}}, \ldots G_{13}^{\text{(3d)}}$.
Nodes may occupy very different positions in the same graphlet.
For example, the three leaf nodes (in black) of $G_4$ are symmetric.
and the other node (in white) of $G_4$ exhibits more like a hub.
According to the positions that nodes of a graphlet occupies, the graphlet's nodes are classified into one or more different orbits (i.e., position classes) associated with them.
The values of orbit IDs have no specific meaning, and we set the values of orbit IDs same as~\cite{PrzuljBioinformatics07}.}
\label{fig:graphlets}
\end{figure*}

Exploring connected and induced subgraph (CIS) patterns (i.e., motifs, also known as graphlets) in a graph is important for understanding and exploring networks such as online social networks (OSNs) and computer networks.
As shown in Fig.~\ref{fig:graphlets}, there is one 2-node undirected graphlet $G_0$, two 3-node undirected graphlets $G_1$ and $G_2$, six 4-node undirected graphlets $G_3, \ldots G_8$,
and thirteen 3-node directed graphlets $G_1^{\text{(3d)}}, \ldots G_{13}^{\text{(3d)}}$, which are widely used for characterizing networks' local connection patterns.
However, nodes may occupy very different positions in the same graphlet.
For example, the three leaf nodes (in black) of $G_4$  in Fig.~\ref{fig:graphlets} are symmetric,
so their positions belong to the same class.
The other node (in white) of $G_4$ behaves more like a hub.
According to the positions that nodes of a graphlet occupies, Przulj et al.~\cite{PrzuljBioinformatics07} group the graphlet's nodes into one or more different automorphism \emph{\textbf{orbits}}\footnote{The values of orbit IDs in Fig.~\ref{fig:graphlets} have no specific meaning. We set the values of orbit IDs same as~\cite{PrzuljBioinformatics07}.} (i.e., position classes).
They observe that a node's graphlet orbit degree vector, or graphlet orbit degree signature, which counts the number of CISes that touch the node at a particular orbit,
is a useful metric for representing the node's topology features.
In fact, the graphlet orbit degree signature has been successfully used for protein function prediction~\cite{Milenkovic2008} and cancer gene identification~\cite{Milenkovic2009} by identifying groups (or clusters) of topologically similar nodes in biological networks.
In addition to biological networks, graphlet orbit degree is also used for link prediction~\cite{YeWWW13} and node classification~\cite{FangTKDE2015} in online social networks, and hyponym relation extraction from Wikipedia hyperlinks~\cite{WeiTKDE2014}.

However, it is computationally intensive to enumerate and compute graphlet orbit degrees for large graphs due to the combinatorial explosion of the problem.
To solve this challenge, approximate methods such as sampling could be used in place of the brute-force enumeration approach.
Despite recent progress in counting specific graphlets such as triangles~\cite{TsourakakisKDD2009,PavanyVLDB2013,JhaKDD2013,AhmedKDD2014} and 4-node motifs~\cite{JhaWWW2015} that appear in a large graph,
little attention has been given to developing fast tools for computing graphlet orbit degrees.
Existing methods of estimating global graphlet counts are customized to sample all CISes in a large graph,
but not tailored to meet the need of sampling CISes that include \textbf{a given node}.

\begin{figure}[htb]
\center
\includegraphics[width=0.49\textwidth]{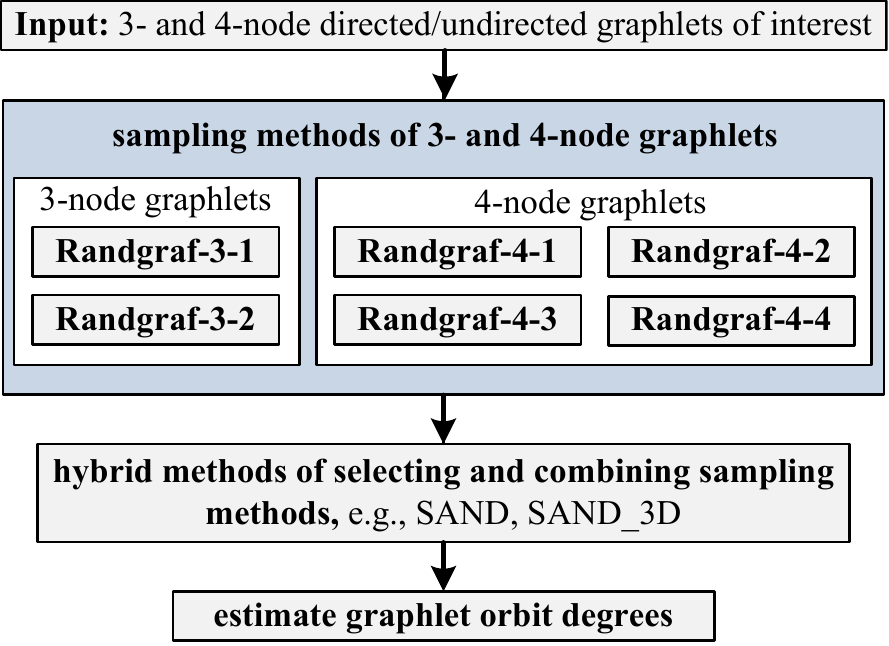}
\caption{Overview of our methods.}
\label{fig:methodoverview}
\end{figure}

To solve this problem, we propose a new method to estimate graphlet orbit degrees and to detect orbits with the largest graphlet orbit degrees for large graphs.
The overview of our method is shown in Fig.~\ref{fig:methodoverview}.
Our contributions are summarized as:

1) We propose a series of methods: Randgraf-3-1, Randgraf-3-2, Randgraf-4-1, Randgraf-4-2, Randgraf-4-3, and Randgraf-4-4 for randomly sampling 3 and 4-node CISes that include a given node.

2) Based on the series of sampling methods, we design scalable and computationally efficient methods, SAND and SAND-3D, to estimate graphlet orbit degrees for undirected and directed graphs respectively, and we also derive expressions for the variances of our estimates, which is of great value in practice since the variances can be used to bound the estimates' errors and determine the smallest necessary sampling budget for a desired accuracy.

3) We conduct experiments on a variety of publicly available datasets.
 Our experimental results show that SAND and SAND-3D are several orders of magnitude faster than state-of-the-art enumeration methods
for accurately estimating graphlet orbit degrees.
We demonstrate the ability of SAND and SAND-3D to explore large graphs with millions of nodes and edges.
To guarantee reproducibility of the experimental results,
we release the source code of SAND in open source\footnote{http://nskeylab.xjtu.edu.cn/dataset/phwang/code}.

The rest of this paper is organized as follows.
Section~\ref{sec:problem} presents the problem formulation.
Section~\ref{sec:preliminaries} introduces preliminaries used in this paper.
Section~\ref{sec:samplingmethods} presents our methods (i.e., Randgraf-3-1, Randgraf-3-2, Randgraf-4-1, Randgraf-4-2, Randgraf-4-3, and Randgraf-4-4) for sampling 3- and 4-node CISes including a given node.
Sections~\ref{sec:undirectedgraphestimation} and~\ref{sec:directedgraphestimation}
present our methods SAND and SAND-3D for estimating undirected and directed graphlet orbit degrees respectively.
Section~\ref{sec:results} presents the performance evaluation and testing results.
Section~\ref{sec:related} summarizes related work. Concluding remarks then follow.

\section{Problem Formulation} \label{sec:problem}
Denote the underlying graph of interest as $G=(V, E, L)$, where $V$ is a set of nodes,
$E$ is a set of \emph{\textbf{undirected}} edges, $E\in V\times V$,
and $L$ is a set of edge directions $\{l_{u,v}: (u,v)\in E\}$,
where  we attach a label
$l_{u,v}\in \{\to, \leftarrow, \leftrightarrow\}$ to
indicate the direction of $(u,v)\in E$ for a directed network.
If $L$ is empty, then $G$ is an undirected graph.

\begin{figure}[htb]
\center
\includegraphics[width=0.4\textwidth]{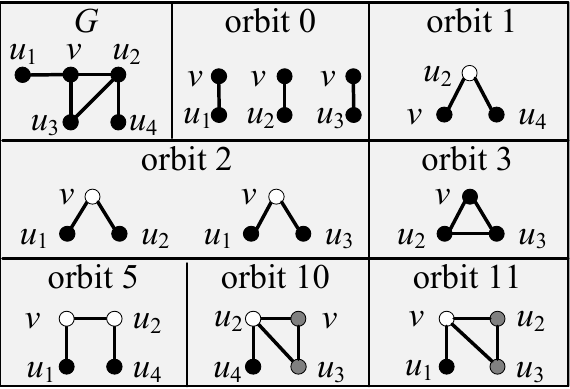}
\caption{Example of computing the undirected orbit degrees of node $v$ in the undirected graph $G$.}
\label{fig:GDDexample}
\end{figure}

In order to define graphlet orbit degrees, we first introduce some notation.
A subgraph $G'$ of $G$ is a graph whose set of nodes, set of edges, and set of edge directions are all subsets of $G$.
An induced subgraph of $G$, $G'=(V', E', L')$, is a subgraph that consists of a subset of nodes in $G$ and \emph{\textbf{all of the edges}} that connect them in $G$,
i.e. $V'\subset V$, $E'=\{(u, v): u, v \in V', (u,v)\in E\}$, $L'=\{l_{u,v}: u, v \in V', (u, v)\in E\}$.
\textbf{\emph{Unless we explicitly say "induced" in this paper, a subgraph is not necessarily induced}.}
Fig.~\ref{fig:GDDgraphlets} shows all
2-, 3-, and 4-node undirected graphlets $G_i$, $0\le i\le 8$, in~\cite{PrzuljBioinformatics07}.
By taking into account the ``symmetries" between nodes in $G_i$,
\cite{PrzuljBioinformatics07} classifies the nodes of $G_i$ into different automorphism \textbf{\emph{orbits}} (or just obits, for brevity),
where the nodes with the same orbit ID are topologically identical.
For all $G_i$, $0\le i\le 8$,
there are 15 orbits, which are shown in Fig.~\ref{fig:GDDgraphlets}.
Denote $C^{(i)}_v$ as the set of connected and induced subgraphs (CISes) in $G$ that touch a node $v\in V$ at orbit $i$.
Let $d^{(i)}_v = |C^{(i)}_v|$ denote the \emph{\textbf{graphlet orbit $i$ degree}} (or just "\emph{\textbf{orbit $i$ degree}}", for brevity) of $v$.
The graphlet orbit degree vector, $(d_v^{(0)}, \ldots, d_v^{(14)})$, can be used as a signature of node $v$ for applications such as identifying similar nodes.
We observe that $C_v^{(0)}$ contains the edges in $G$ that includes node $v$, i.e., $C_v^{(0)} = \{(u,v): (u,v)\in E\}$,
and $d_v^{(0)}$ is the number of neighbors of $v$.
For simplicity, we denote $d_v = d_v^{(0)}$ as the degree of node $v$.
An example is given in Fig.~\ref{fig:GDDexample},
where $d^{(0)}_v = 3$, $d^{(2)}_v = 2$, $d^{(1)}_v = d^{(3)}_v = d^{(5)}_v = d^{(10)}_v = d^{(11)}_v = 1$, and $ d^{(4)}_v = d^{(6)}_v =d^{(7)}_v = d^{(8)}_v = d^{(9)}_v = d^{(12)}_v = d^{(13)}_v = d^{(14)}_v = 0$.
The concept of orbit and graphlet orbit degree extends to directed graphs,
As shown in Fig.~\ref{fig:othergraphlets}, directed graphs have thirteen 3-node graphlets $G_1^{(3\text{d})}$ whose nodes are distributed at 30 different orbits.
In this paper we focus on 3-node directed graphlets and orbits because of the large number of directed 4-node graphlets and orbits.

As discussed above, it is computationally intensive to enumerate and count all 3- and 4-node CISes that include a given node with a large number of neighbors in large graphs.
For example, later our experiments show that the node with the largest degree in graph Wiki-Talk~\cite{Leskovec2010} belongs to more than $10^{14}$ 3- and 4-node CISes.
In this paper, we develop efficient methods to estimate graphlet orbit degrees and identify orbits with the largest graphlet orbit degrees for undirected and directed graphs.
For ease of reading, we list notation used throughout the paper in Table~\ref{tab:notations}
and we present the proofs of all our theorems in Appendix.

\begin{table}[htb]
\begin{center}
\caption{Table of notation.\label{tab:notations}}
\begin{tabular}{|c|l|} \hline
$G=(V, E, L)$&$G$ is the graph of interest\\ \hline
$N_v$&the set of neighbors of a node $v$ in $G$ \\ \hline
$d_v$& $d_v = |N_v|$, the cardinality of set $N_v$\\ \hline
$G_0, \ldots, G_8$ &2-, 3-, and 4-node undirected graphlets \\\hline
$G_1^{3\text{d}}, \ldots, G_{13}^{3\text{d}}$ &3-node directed graphlets \\\hline
$d^{(1)}_v, \ldots, d^{(14)}_v$ &undirected orbit degrees of node $v$ \\\hline
$d^{(1,\text{dir})}_v, \ldots, d^{(30,\text{dir})}_v$ &directed orbit degrees of node $v$ \\\hline
$p_1^{(3,1)}, \ldots, p_{14}^{(3,1)}$&probability distribution of methods\\
$p_1^{(3,2)}, \ldots, p_{14}^{(3,2)}$&Randgraf-3-1, Randgraf-3-2,\\
$p_1^{(4,1)}, \ldots, p_{14}^{(4,1)}$&Randgraf-4-1, Randgraf-4-2,\\
$p_1^{(4,2)}, \ldots, p_{14}^{(4,2)}$&Randgraf-4-3, and Randgraf-4-4\\
$p_1^{(4,3)}, \ldots, p_{14}^{(4,3)}$&sampling undirected orbits 1--14\\
$p_1^{(4,4)}, \ldots, p_{14}^{(4,4)}$&respectively\\\hline
\tabincell{c}{$K^{(3,1)}$, $K^{(3,2)}$,\\$K^{(4,1)}$, $K^{(4,2)}$,\\$K^{(4,3)}$, $K^{(4,4)}$}
&\tabincell{l}{sampling budgets of Randgraf-3-1,\\Randgraf-3-2, Randgraf-4-1,\\Randgraf-4-2, Randgraf-4-3,\\and Randgraf-4-4}\\\hline
\multicolumn{2}{|c|}{$\alpha^{(v)}=\{\alpha_u^{(v)} = \frac{d_u-1}{\varphi_v }: u\in N_v\}$}\\ \hline
\multicolumn{2}{|c|}{$\beta^{(v)} = \{\beta_u^{(v)} = \frac{\phi_u - d_u + 1}{\Phi_v^{(2)}}: u\in N_v\}$}\\ \hline
\multicolumn{2}{|c|}{$\gamma^{(v)} = \{\gamma_u^{(v)} = \frac{\varphi_u - d_v + 1}{\Phi_v^{(3)}}: u\in N_v\}$} \\ \hline
\multicolumn{2}{|c|}{$\rho^{(u, v)} = \{\rho_w^{(u, v)} = \frac{d_w - 1}{\varphi_u - d_v + 1}: w\in N_u-\{v\}\}$}\\ \hline
\multicolumn{2}{|c|}{$\phi_v = \frac{d_v (d_v - 1)}{2}, \quad \varphi_v = \sum_{u\in N_v} (d_u - 1)$ } \\ \hline
\multicolumn{2}{|c|}{$\Phi_v^{(1)}= (d_v - 1) \varphi_v, \quad \Phi_v^{(2)} = \sum_{u\in N_v} (\phi_u - d_u + 1)$} \\ \hline
\multicolumn{2}{|c|}{$\Phi_v^{(3)}= \sum_{u\in N_v} (\varphi_u - d_v + 1), \quad \Phi_v^{(4)} = \frac{d_v (d_v - 1)(d_v - 2)}{6}$} \\ \hline
\end{tabular}
\end{center}
\end{table}

\section{Preliminaries}\label{sec:preliminaries}
In this section, we introduce two theorems
that provide the foundation for our methods of estimating graphlet orbit degrees.
\begin{theorem}\label{theorem:estimatecardinality}
(\textbf{Estimating subset cardinalities})
Let $S_1, \ldots, S_r$ be a non-overlapping division of a set $S$ of interest,
i.e., $S=S_1\cup \ldots \cup S_r$ and $S_i\cap S_j = \emptyset$, $i\ne j$, $i, j = 1, \ldots, r$.
Let $n_i = |S_i|$ denote the cardinality of $S_i$, $1\le i\le r$.
Suppose there exists a function $\mathbb{F}$ that returns an item $X$ sampled from $S$ according to a distribution $P(X=s, s\in S_i)=p_i$, where $\sum_{i=1}^r n_i p_i = 1$.
Let $X_1, \ldots, X_K$ be items obtained by calling function $\mathbb{F}$ $K$ times independently.
Denote by $\mathbf{1}(\mathbb{X})$ the indicator function that equals one when predicate $\mathbb{X}$ is true, and zero otherwise.
When $p_i>0$, we can estimate $n_i$ as:
\[
\hat n_i = \frac{\sum_{j=1}^K \mathbf{1}(X_j\in S_i)}{K p_i}, \quad 1\le i\le r,
\]
where $\hat n_i$ is an \textbf{unbiased} estimator of $n_i$, i.e., $\mathbb{E}(\hat n_i) = n_i$
with variance $\text{Var}(\hat n_i) = \frac{n_i}{K}\left(\frac{1}{p_i} - n_i\right)$.
The covariance of $\hat n_i$ and $\hat n_j$ is
$\text{Cov}(\hat n_i, \hat n_j) = -\frac{n_i n_j}{K}$, $i\ne j, i,j=1, \ldots, r$.
\end{theorem}

\begin{theorem}\label{theorem:mixestimators}
(\textbf{Combining unbiased estimators}\cite{Franklin1959})
Suppose there exist $k$ \textbf{independent} and \textbf{unbiased} estimates $c_1, \ldots, c_k$ of $c$
with variances $\text{Var}(c_j)$, $j=1,\ldots,k$.
The estimate $\hat c = \sum_{j=1}^k \alpha_j c_j$ where $\alpha_i = \frac{\text{Var}^{-1}(c_i)}{\sum_{j=1}^k \text{Var}^{-1}(c_j)}$ is the minimum variance estimate based on a linear combination of $c_1, \ldots, c_k$.
It has variance $\text{Var}(\hat c) = \frac{1}{\sum_{j=1}^k \text{Var}^{-1}(c_j)}$.
\end{theorem}

\section{Sampling 3- and 4-Node CISes}\label{sec:samplingmethods}
In this section, we first present the basic idea and then present methods for sampling 3- and 4-node CISes.
\textbf{Basic idea behind our methods}:
Let $S(v)$ denote the set of all 3- and 4-node CISes that include a  given node $v\in V$ of interest.
Let $S_i(v)\in S(v)$ denote the set of CISes that include $v$ in undirected orbit $i=1,\ldots, 14$ (Fig.~\ref{fig:GDDgraphlets}).
According to Theorem~\ref{theorem:estimatecardinality},
the key to estimating orbit degrees of $v$ is to design a fast method to sample CISes from $S(v)$ whose sampling probability distribution $P(X=s, s\in S_i)$, $1\le i\le r$, can be easily derived and computed.
Our sampling methods are performed on the undirected graphs of $G$.
The "orbit" mentioned in this section refers to the "undirected orbit".
These sampling methods are used as building blocks for graphlet statistics estimation methods presented in Sections~\ref{sec:undirectedgraphestimation} and~\ref{sec:directedgraphestimation}.

\subsection{Methods for Sampling 3-Node CISes}\label{sec: 3-node-sampling}
We develop two efficient sampling methods $\text{Randgraf-3-1}(v, G)$ and $\text{Randgraf-3-2}(v, G)$ to sample 3-node CISes in $G$ that include $v$.
$\text{Randgraf-3-1}(v, G)$ is able to sample 3-node CISes that include $v$ in orbits 2 and 3.
$\text{Randgraf-3-2}(v, G)$ is able to sample 3-node CISes that include $v$ in orbits 1 and 3.
Next, we introduce $\text{Randgraf-3-1}(v, G)$ and $\text{Randgraf-3-2}(v, G)$ respectively.

\textbf{Method} $\textbf{Randgraf-3-1}({v, G})$: To sample a CIS that includes $v$,
$\text{Randgraf-3-1}(v, G)$ consists of three steps:
\textbf{Step 1}) Sample node $u$ from $N_v$ (i.e., the neighbors of $v$) at random;
\textbf{Step 2}) Sample node $w$ from $N_v\setminus \{u\}$ at random;
\textbf{Step 3}) Return CIS $s$ consisting of nodes $v$, $u$, and $w$.
The pseudo-code for $\text{Randgraf-3-1}(v, G)$ is shown in Algorithm~\ref{alg:randgraf-3-1}.
Theorem~\ref{theorem:prob_randgraf-3-1} specifies the sampling bias of $\text{Randgraf-3-1}(v, G)$, which is critical for estimating graphlet orbit degrees of $v$.
\begin{algorithm}
\SetKwRepeat{Do}{do}{while}%
\SetKwFunction{CIS}{CIS}
\SetKwFunction{WeightRandomVertex}{WeightRandomVertex}
\SetKwFunction{RandomVertex}{RandomVertex}
\SetKwInOut{Input}{input}
\SetKwInOut{Output}{output}
\Input{$G=(V, E, L)$ and $v\in V$.}
\Output{a 3-node CIS $s$ that includes $v$.}
$u \gets \RandomVertex(N_v)$\;
$w \gets \RandomVertex(N_v\setminus \{u\})$\;
$s\gets  \CIS(\{v,u,w\})$\;
\caption{The pseudo-code of $\text{Randgraf-3-1}(v, G)$. \label{alg:randgraf-3-1}}
\end{algorithm}

\begin{theorem}\label{theorem:prob_randgraf-3-1}
Let $p_i^{(3,1)}$, $i\in \{1,2,3\}$, denote the probability that method Randgraf-3-1 samples a 3-node CIS $s$ including $v$ in orbit $i$.
Then $p_1^{(3,1)} = 0$, $p_2^{(3,1)}=\frac{1}{\phi_v}$, and $p_3^{(3,1)}=\frac{1}{\phi_v}$, where $\phi_v = \frac{d_v (d_v - 1)}{2}$.
\end{theorem}

\textbf{Method} $\textbf{Randgraf-3-2}({v, G})$: Define $\varphi_v = \sum_{u\in N_v} (d_u - 1)$ and $\alpha_u^{(v)} = \frac{d_u-1}{\varphi_v }$.
To sample a 3-node CIS that includes $v$,
method $\text{Randgraf-3-2}(v, G)$ consists of three steps:
\textbf{Step 1}) Sample node $u$ from $N_v$ according to distribution $\alpha^{(v)}=\{\alpha_u^{(v)}: u\in N_v\}$.
Here we do not sample $u$ from $N_v$ uniformly but according to $\alpha^{(v)}$
to facilitate estimation of the sampling bias;
\textbf{Step 2}) Sample node $w$ from $N_u\setminus \{v\}$ at random;
\textbf{Step 3}) Return CIS $s$ consisting of nodes $v$, $u$, and $w$.
Algorithm~\ref{alg:randgraf-3-2} shows the pseudo-code for $\text{Randgraf-3-2}(v, G)$.
Function $\text{WeightRandomVertex}(N_v, \alpha^{(v)})$  in Algorithm~\ref{alg:randgraf-3-2} returns a node sampled from $N_v$ according to distribution $\alpha^{(v)}=\{\alpha_u^{(v)}: u\in N_v\}$.
Theorem~\ref{theorem:prob_randgraf-3-2} specifies the sampling bias of $\text{Randgraf-3-2}(v, G)$.

\begin{algorithm}
\SetKwFunction{CIS}{CIS}
\SetKwFunction{WeightRandomVertex}{WeightRandomVertex}
\SetKwFunction{RandomVertex}{RandomVertex}
\SetKwInOut{Input}{input}
\SetKwInOut{Output}{output}
\Input{$G=(V, E, L)$ and $v\in V$.}
\Output{a 3-node CIS $s$ that includes $v$.}
$u \gets \WeightRandomVertex(N_v, \alpha^{(v)})$\;
$w \gets \RandomVertex(N_u\setminus \{v\})$\;
$s\gets  \CIS(\{v,u,w\})$\;
\caption{The pseudo-code of $\text{Randgraf-3-2}(v, G)$. \label{alg:randgraf-3-2}}
\end{algorithm}

\begin{theorem}\label{theorem:prob_randgraf-3-2}
Let $p_i^{(3,2)}$, $i\in \{1,2,3\}$, denote the probability that method Randgraf-3-2 samples a 3-node CIS $s$ including $v$ in orbit $i$.
Then $p_1^{(3,2)} = \frac{1}{\varphi_v}$, $p_2^{(3,2)}=0$, and $p_3^{(3,2)}=\frac{2}{\varphi_v}$.
\end{theorem}

\subsection{Methods for Sampling 4-Node CISes}\label{sec: 4-node-sampling}
In this subsection, we develop four methods: $\text{Randgraf-4-1}(v, G)$, $\text{Randgraf-4-2}(v, G)$, $\text{Randgraf-4-3}(v, G)$, and $\text{Randgraf-4-4}(v, G)$ to sample 4-node CISes in $G$ that include $v$.
Each of these four methods is only able to sample 4-node CISes that include $v$ in \textbf{a subset of orbits}.
However, together they are able to sample all 4-node CISes that include $v$,
We introduce these four methods below.

\textbf{Method} $\textbf{Randgraf-4-1}({v, G})$:
To sample a 4-node CIS that includes $v$,
method $\text{Randgraf-4-1}(v, G)$ consists of four steps:
\textbf{Step 1}) Sample node $u$ from $N_v$ according to distribution $\alpha^{(v)} = \{\alpha_u^{(v)}: u\in N_v\}$;
\textbf{Step 2}) Sample node $w$ from $N_v \setminus\{u\}$ at random;
\textbf{Step 3}) Sample node $r$ from $N_u \setminus\{v\}$ at random;
\textbf{Step 4}) Return CIS $s$ consisting of nodes $v$, $u$, $w$, and $r$.
Note that $s$ is a 3-node CIS when $w = r$.
The pseudo-code of $\text{Randgraf-4-1}(v, G)$ is shown in Algorithm~\ref{alg:randgraf-4-1}.
Theorem~\ref{theorem:prob_randgraf-4-1} states the sampling bias of $\text{Randgraf-4-1}(v, G)$, where $\Phi_v^{(1)}= (d_v - 1) \varphi_v$.

\begin{algorithm}
\SetKwFunction{CIS}{CIS}
\SetKwFunction{WeightRandomVertex}{WeightRandomVertex}
\SetKwFunction{RandomVertex}{RandomVertex}
\SetKwInOut{Input}{input}
\SetKwInOut{Output}{output}
\Input{$G=(V, E, L)$ and $v\in V$.}
\Output{a 3- or 4-node CIS $s$ that includes $v$.}
$u \gets \WeightRandomVertex(N_v, \alpha^{(v)})$\;
$w \gets \RandomVertex(N_v\setminus \{u\})$\;
$r \gets \RandomVertex(N_u\setminus \{v\})$\;
$s \gets  \CIS(\{v, u, w, r\})$\;
\caption{The pseudo-code of $\text{Randgraf-4-1}(v, G)$. \label{alg:randgraf-4-1}}
\end{algorithm}

\begin{theorem}\label{theorem:prob_randgraf-4-1}
Let $p_i^{(4,1)}$, $i\in \{1,\ldots,14\}$, denote the probability that method Randgraf-4-1 samples a 3- or 4-node CIS $s$ including $v$ in orbit $i$.
Then $p_1^{(4,1)} = p_2^{(4,1)} = p_4^{(4,1)} = p_6^{(4,1)} = p_7^{(4,1)} = p_9^{(4,1)} = 0$,
$p_3^{(4,1)}=\frac{2}{\Phi_v^{(1)}}$,
$p_5^{(4,1)}=\frac{1}{\Phi_v^{(1)}}$,
$p_8^{(4,1)}=\frac{2}{\Phi_v^{(1)}}$,
$p_{10}^{(4,1)}=\frac{1}{\Phi_v^{(1)}}$,
$p_{11}^{(4,1)}=\frac{2}{\Phi_v^{(1)}}$,
$p_{12}^{(4,1)}=\frac{2}{\Phi_v^{(1)}}$,
$p_{13}^{(4,1)}=\frac{4}{\Phi_v^{(1)}}$,
and $p_{14}^{(4,1)}=\frac{6}{\Phi_v^{(1)}}$.
\end{theorem}

\textbf{Method} $\textbf{Randgraf-4-2}({v, G})$:
Define $\Phi_v^{(2)} = \sum_{u\in N_v} (\phi_u - d_u + 1)$ and $\beta_u^{(v)} = \frac{\phi_u - d_u + 1}{\Phi_v^{(2)}}$.
To sample a 4-node CIS that includes $v$,
$\text{Randgraf-4-2}(v, G)$ consists of four steps:
\textbf{Step 1}) Sample node $u$ from $N_v$ according to distribution $\beta^{(v)} = \{\beta_u^{(v)}: u\in N_v\}$;
\textbf{Step 2}) Sample node $w$ from $N_u\setminus \{v\}$ at random;
\textbf{Step 3}) Sample node $r$ from $N_u\setminus \{v, u\}$ at random;
\textbf{Step 4}) Return CIS $s$ consisting of nodes $v$, $u$, $w$, and $r$.
Theorem~\ref{theorem:prob_randgraf-4-2} states the sampling bias of $\text{Randgraf-4-2}(v, G)$.

\begin{algorithm}
\SetKwFunction{CIS}{CIS}
\SetKwFunction{WeightRandomVertex}{WeightRandomVertex}
\SetKwFunction{RandomVertex}{RandomVertex}
\SetKwInOut{Input}{input}
\SetKwInOut{Output}{output}
\Input{$G=(V, E, L)$ and $v\in V$.}
\Output{a 4-node CIS $s$ that includes $v$.}
$u \gets \WeightRandomVertex(N_v, \beta^{(v)})$\;
$w \gets \RandomVertex(N_u\setminus \{v\})$\;
$r \gets \RandomVertex(N_u\setminus \{v, u\})$\;
$s \gets  \CIS(\{v, u, w, r\})$\;
\caption{The pseudo-code of $\text{Randgraf-4-2}(v, G)$. \label{alg:randgraf-4-2}}
\end{algorithm}

\begin{theorem}\label{theorem:prob_randgraf-4-2}
Let $p_i^{(4,2)}$, $i\in \{1,\ldots,14\}$, denote the probability that method Randgraf-4-2 samples a 4-node CIS $s$ including $v$ in orbit $i$.
Then $p_1^{(4,2)} = p_2^{(4,2)} = p_3^{(4,2)} = p_4^{(4,2)} = p_5^{(4,2)} = p_7^{(4,2)} = p_8^{(4,2)} = p_{11}^{(4,2)} = 0$,
$p_6^{(4,2)}=\frac{1}{\Phi_v^{(2)}}$,
$p_9^{(4,2)}=\frac{1}{\Phi_v^{(2)}}$,
$p_{10}^{(4,2)}=\frac{1}{\Phi_v^{(2)}}$,
$p_{12}^{(4,2)}=\frac{2}{\Phi_v^{(2)}}$,
$p_{13}^{(4,2)}=\frac{1}{\Phi_v^{(2)}}$,
and $p_{14}^{(4,2)}=\frac{3}{\Phi_v^{(2)}}$.
\end{theorem}

\textbf{Method} $\textbf{Randgraf-4-3}({v, G})$: Define $\Phi_v^{(3)}= \sum_{u\in N_v} (\varphi_u - d_v + 1)$, $\gamma_u^{(v)} = \frac{\varphi_u - d_v + 1}{\Phi_v^{(3)}}$,
and $\rho_w^{(u, v)} = \frac{d_w - 1}{\varphi_u - d_v + 1}$.
To sample a 4-node CIS that includes $v$,
method $\text{Randgraf-4-3}(v, G)$ consists of four steps:
\textbf{Step 1}) Sample node $u$ from $N_v$ according to distribution $\gamma^{(v)} = \{\gamma_u^{(v)}: u\in N_v\}$;
\textbf{Step 2}) Sample node $w$ from $N_u\setminus \{v\}$ according to distribution $\rho^{(u, v)} = \{\rho_w^{(u, v)}: w\in N_u\setminus \{v\}\}$;
\textbf{Step 3}) Sample node $r$ from $N_w\setminus \{u\}$ at random;
\textbf{Step 4}) Return CIS $s$ consisting of nodes $v$, $u$, $w$, and $r$.
Note that $s$ is a 3-node CIS when $r = v$.
The pseudo-code for $\text{Randgraf-4-3}(v, G)$ is shown in Algorithm~\ref{alg:randgraf-4-3}.
Theorem~\ref{theorem:prob_randgraf-4-3} states the sampling bias of $\text{Randgraf-4-3}(v, G)$.

\begin{algorithm}
\SetKwFunction{CIS}{CIS}
\SetKwFunction{WeightRandomVertex}{WeightRandomVertex}
\SetKwFunction{RandomVertex}{RandomVertex}
\SetKwInOut{Input}{input}
\SetKwInOut{Output}{output}
\Input{$G=(V, E, L)$ and $v\in V$.}
\Output{a 3- or 4-node CIS $s$ that includes $v$.}
$u \gets \WeightRandomVertex(N_v, \gamma^{(v)})$\;
$w \gets \WeightRandomVertex(N_u\setminus \{v\}, \rho^{(u, v)})$\;
$r \gets \RandomVertex(N_w\setminus \{u\})$\;
$s \gets  \CIS(\{v,u,w,r\})$\;
\caption{The pseudo-code of $\text{Randgraf-4-3}(v, G)$. \label{alg:randgraf-4-3}}
\end{algorithm}

\begin{theorem}\label{theorem:prob_randgraf-4-3}
Let $p_i^{(4,3)}$, $i\in \{1,\ldots,14\}$, denote the probability that method Randgraf-4-3 samples a 3- or 4-node CIS $s$ including $v$ in orbit $i$.
Then $p_1^{(4,3)} = p_2^{(4,3)} = p_5^{(4,3)} = p_6^{(4,3)} = p_7^{(4,3)} = p_{11}^{(4,3)} = 0$,
$p_3^{(4,3)}=\frac{2}{\Phi_v^{(3)}}$,
$p_4^{(4,3)}=\frac{1}{\Phi_v^{(3)}}$,
$p_8^{(4,3)}=\frac{2}{\Phi_v^{(3)}}$,
$p_9^{(4,3)}=\frac{2}{\Phi_v^{(3)}}$,
$p_{10}^{(4,3)}=\frac{1}{\Phi_v^{(3)}}$,
$p_{12}^{(4,3)}=\frac{4}{\Phi_v^{(3)}}$,
$p_{13}^{(4,3)}=\frac{2}{\Phi_v^{(3)}}$,
and $p_{14}^{(4,3)}=\frac{6}{\Phi_v^{(3)}}$.
\end{theorem}

\textbf{Method} $\textbf{Randgraf-4-4}({v, G})$: To sample a 4-node CIS that includes $v$,
method $\text{Randgraf-4-4}(v, G)$ consists of four steps:
\textbf{Step 1}) Sample node $u$ from $N_v$ at random;
\textbf{Step 2}) Sample node $w$ from $N_v\setminus \{u\}$ at random;
\textbf{Step 3}) Sample node $r$ from $N_v\setminus \{v, u\}$ at random;
\textbf{Step 4}) Return CIS $s$ consisting of nodes $v$, $u$, $w$, and $r$.
The pseudo-code for $\text{Randgraf-4-4}(v, G)$ is shown in Algorithm~\ref{alg:randgraf-4-4}.
Theorem~\ref{theorem:prob_randgraf-4-4} states the sampling bias of $\text{Randgraf-4-4}(v, G)$, where $\Phi_v^{(4)} = \frac{d_v (d_v - 1)(d_v - 2)}{6}$.

\begin{algorithm}
\SetKwRepeat{Do}{do}{while}%
\SetKwFunction{CIS}{CIS}
\SetKwFunction{WeightRandomVertex}{WeightRandomVertex}
\SetKwFunction{RandomVertex}{RandomVertex}
\SetKwInOut{Input}{input}
\SetKwInOut{Output}{output}
\Input{$G=(V, E, L)$ and $v\in V$ with $d_v\ge 3$.}
\Output{a 4-node CIS $s$ that includes $v$.}
$u \gets \RandomVertex(N_v)$\;
$w \gets \RandomVertex(N_v\setminus \{u\})$\;
$r \gets \RandomVertex(N_v\setminus \{u, v\})$\;
$s \gets  \CIS(\{v,u,w,r\})$\;
\caption{The pseudo-code of $\text{Randgraf-4-4}(v, G)$. \label{alg:randgraf-4-4}}
\end{algorithm}

\begin{theorem}\label{theorem:prob_randgraf-4-4}
Let $p_i^{(4,4)}$, $i\in \{1,\ldots,14\}$, denote the probability that method Randgraf-4-4 samples a 3- or 4-node CIS $s$ including $v$ in orbit $i$.
Then $p_1^{(4,4)} = p_2^{(4,4)} = p_3^{(4,4)} = p_4^{(4,4)} = p_5^{(4,4)} = p_6^{(4,4)} = p_8^{(4,4)} = p_9^{(4,4)} = p_{10}^{(4,4)} = p_{12}^{(4,4)} = 0$, and $p_{7}^{(4,4)}=p_{11}^{(4,4)}=p_{13}^{(4,4)} = p_{14}^{(4,4)}=\frac{1}{\Phi_v^{(4)}}$.
\end{theorem}

\subsection{Discussion}
The details of implementing the functions in the Algorithms we presented subsections~\ref{sec: 3-node-sampling} and~\ref{sec: 4-node-sampling}
and analyzing their computational complexities are discussed in Appendix.
Table~\ref{tab:samplingmethods} summarizes and compares Randgraf-3-1, Randgraf-3-2, Randgraf-4-1, Randgraf-4-2, Randgraf-4-3, and Randgraf-4-4.
We observe that
1) each method is not able to sample "\textbf{all}" CISes that includes $v$. CISes that includes $v$ in orbits 2, 1, 5, 6, 4, and 7 can only be sampled by Randgraf-3-1, Randgraf-3-2, Randgraf-4-1, Randgraf-4-2, Randgraf-4-3, and Randgraf-4-4 respectively.
Thus, all six methods are needed to guarantee each CIS that includes $v$ is sampled with probability larger than zero;
2) Randgraf-4-1, Randgraf-4-2, and Randgraf-4-3 are able to sample CISes that includes $v$ in more orbits than Randgraf-4-4;
%3) Randgraf-3-1 exhibits lower computational complexity than Randgraf-3-2;
3) Randgraf-4-3 exhibits the highest computational complexity than the other methods.

\begin{table*}[htb]
\begin{center}
\caption{Summary of graphlet orbit sampling methods in this paper.\label{tab:samplingmethods}}
\begin{tabular}{|c|c|c|c|c|c|c|c|c|c|c|c|c|c|c|c|c|}\hline
\multirow{2}{*}{\textbf{method}}&\multicolumn{14}{c|}{\textbf{whether the method is able to sample a CIS that includes $v$ in orbit $i$}}&\multicolumn{2}{c|}{\textbf{computational complexity}}\\
\cline{2-17}
&\textbf{1}&\textbf{2}&\textbf{3}&\textbf{4}&\textbf{5}&\textbf{6}&\textbf{7}&\textbf{8}&\textbf{9}&\textbf{10}&\textbf{11}&\textbf{12}&\textbf{13}&\textbf{14}&\textbf{initialization}&\textbf{sample one CIS}\\ \hline
$\text{Randgraf-3-1}(v,G)$&$\times$&{\color{red}$\checkmark$}&$\checkmark$&$\times$&$\times$&$\times$&$\times$&$\times$&$\times$&$\times$&$\times$&$\times$&$\times$&$\times$&0&$O(1)$\\ \hline
$\text{Randgraf-3-2}(v,G)$&{\color{red}$\checkmark$}&$\times$&$\checkmark$&$\times$&$\times$&$\times$&$\times$&$\times$&$\times$&$\times$&$\times$&$\times$&$\times$&$\times$&0&$O(\log d_v)$\\ \hline
$\text{Randgraf-4-1}(v,G)$&$\times$&$\times$&$\checkmark$&$\times$&{\color{red}$\checkmark$}&$\times$&$\times$&$\checkmark$&$\times$&$\checkmark$&$\checkmark$&$\checkmark$&$\checkmark$&$\checkmark$&$O(d_v)$&$O(\log d_v)$\\ \hline
$\text{Randgraf-4-2}(v,G)$&$\times$&$\times$&$\times$&$\times$&$\times$&{\color{red}$\checkmark$}&$\times$&$\times$&$\checkmark$&$\checkmark$&$\times$&$\checkmark$&$\checkmark$&$\checkmark$&$O(d_v)$&$O(\log d_v)$\\ \hline
\multirow{2}{*}{$\text{Randgraf-4-3}(v,G)$}&\multirow{2}{*}{$\times$}&\multirow{2}{*}{$\times$}&\multirow{2}{*}{$\checkmark$}&\multirow{2}{*}{{\color{red}$\checkmark$}}&\multirow{2}{*}{$\times$}&\multirow{2}{*}{$\times$}&\multirow{2}{*}{$\times$}&\multirow{2}{*}{$\checkmark$}&\multirow{2}{*}{$\checkmark$}&\multirow{2}{*}{$\checkmark$}&\multirow{2}{*}{$\times$}&\multirow{2}{*}{$\checkmark$}&\multirow{2}{*}{$\checkmark$}&\multirow{2}{*}{$\checkmark$}&$O(d_v +$&$O(\log d_v +$\\
&&&&&&&&&&&&&&&$\sum_{u\in N_v} d_u)$&$\sum_{u\in N_v} \pi_u^{(v)} \log d_u)$\\ \hline
$\text{Randgraf-4-4}(v,G)$&$\times$&$\times$&$\times$&$\times$&$\times$&$\times$&{\color{red}$\checkmark$}&$\times$&$\times$&$\times$&$\checkmark$&$\times$&$\checkmark$&$\checkmark$&0&$O(1)$\\ \hline
\end{tabular}
\end{center}
\end{table*}

\section{SAND: Estimation of Undirected Orbit Degrees}\label{sec:undirectedgraphestimation}
In this section, we present orbit degree estimators based on the above sampling methods.
We first focus on a single undirected orbit and then modify it to estimate the degrees of all undirected orbits.
\subsection{Estimating Single Undirected Orbit Degree}\label{sec:undirectedgraphsingle}
Consider the problem of estimating the orbit $i$ degree.
If undirected orbit $i$ can only be sampled by one method in Section~\ref{sec:preliminaries}
(e.g., Randgraf-3-2 is the only that samples orbit 1),
we use that method to obtain $K$ CISes that include node $v\in V$.
Let $p_i$ be the probability that the method samples a CIS in orbit $i$,
and let $m_i$ denote the number of sampled CISes that include $v$ in orbit $i$.
According to Theorem~\ref{theorem:estimatecardinality},
we estimate $d^{(i)}_v $ as
\begin{equation*}
\hat d^{(i)}_v = \frac{m_i}{K p_i},
\end{equation*}
and the variance of $\hat d^{(i)}_v $ is $\text{Var}(\hat d^{(i)}_v) = \frac{d^{(i)}_v}{K}\left(\frac{1}{p_i} - d^{(i)}_v\right)$.
When more than one method is able to sample undirected orbit $i$,
we select the most efficient method, the one with the smallest $\frac{\text{Var}(\hat d^{(i)}_v)}{K t_v}$ to estimate $d^{(i)}_v$, where $t_v$ is the average computational time of the method sampling a CIS.

\subsection{Estimating all Undirected Orbit Degrees}
We observe that \textbf{the relationships between undirected orbit degrees can be used to reduce the sampling cost of estimating all undirected orbit degrees}.
For example, the following Theorem~\ref{theorem:equation_undirected} show that $d_v^{(2)} + d_v^{(3)} = \phi_v$.
When one has obtained an accurate estimate of $d_v^{(3)}$, 
it is not necessary to apply the sampling method in Section~\ref{sec:undirectedgraphsingle} to estimate $d_v^{(2)}$ 
since $d_v^{(2)}$ can be computed according to the above equation.  

\begin{theorem}\label{theorem:equation_undirected}
We have the following relations for a node $v\in V$ in undirected graph $G$
\begin{equation}\label{eq:randgraf-3-2}
d^{(2)}_v + d^{(3)}_v = \phi_v,
\end{equation}
\begin{equation}\label{eq:randgraf-4-3}
\begin{split}
&2 d^{(3)}_v + d^{(4)}_v  + 2 d^{(8)}_v + 2 d^{(9)}_v + d^{(10)}_v + 4 d^{(12)}_v + 2 d^{(13)}_v\\
&+ 6 d^{(14)}_v = \Phi_v^{(3)},
\end{split}
\end{equation}
\begin{equation}\label{eq:randgraf-4-4}
d^{(7)}_v + d^{(11)}_v + d^{(13)}_v + d^{(14)}_v = \Phi_v^{(4)}.
\end{equation}
\end{theorem}

We develop a fast method SAND consisting of Randgraf-3-2, Randgraf-4-1, Randgraf-4-2 to estimate all 3- and 4-node undirected orbit degrees
inspired by the following observations:\\
\textbf{Observation 1}. For node $v$ with the largest degree in $G$, we observe $d_v^{(2)}\gg d_v^{(1)}$ and $d_v^{(2)}\gg d_v^{(3)}$ for most real-world networks.
Then, we find that Randgraf-3-2 is more efficient for estimating $d_v^{(3)}$ than Randgraf-3-1
because Randgraf-3-1 rarely samples undirected orbit 3.
Similarly, we observe that Randgraf-4-1, Randgraf-4-2,  Randgraf-4-3, and Randgraf-4-4 never or rarely sample undirected orbit 3.\\
\textbf{Observation 2}. $\text{Randgraf-4-1}(v, G)$ and  $\text{Randgraf-4-2}(v, G)$
together can sample 4-node CISes that include $v$ in undirected orbits $i\in \{4, \ldots, 14\}\setminus \{ 4, 7\}$.\\
\textbf{Observation 3}. Theorem~\ref{theorem:equation_undirected} presents three relationships between undirected orbit degrees,
which enable us to estimate undirect orbit $i\in \{2, 4, 7\}$ degrees.

Formally, SAND consists of the following three steps:\\
\textbf{Step 1}: Apply function $\text{Randgraf-3-2}(v, G)$ $ K^{(3,2)}$ times to sample $ K^{(3,2)}$ CISes,
and then count the number of sampled CISes that include $v$ in undirected orbit $i\in \{1, 2, 3\}$, denoted as $m_i^{(3,2)}$;\\
\textbf{Step 2}: Apply $\text{Randgraf-4-1}(v, G)$ $ K^{(4,1)}$ times to sample $ K^{(4,1)}$ CISes,
and then count the number of sampled CISes that include $v$ in undirected orbit $i\in \{1, \ldots, 14\}$, denoted as $m_i^{(4,1)}$;\\
\textbf{Step 3}: Apply function $\text{Randgraf-4-2}(v, G)$ $ K^{(4,2)}$ times to sample $ K^{(4,2)}$ CISes,
and then count the number of sampled CISes that include $v$ in undirected orbit $i\in \{1, \ldots, 14\}$, denoted as $m_i^{(4,2)}$.

Next, we estimate $d^{(1)}_v, \ldots, d^{(14)}_v$ as follows:\\
\textbf{Step 1}: For undirected orbit $i\in \{1, 5, 6, 8, 9, 11\}$, we estimate $d^{(i)}_v$ as
\begin{equation} \label{eq:hat_d_single}
\begin{split}
\hat{d}^{(i)}_v &=
\begin{dcases}
\frac{m_1^{(3,2)}}{K^{(3,2)} p_2^{(3,2)}},  & i=1,\\
\frac{m_i^{(4,1)}}{K^{(4,1)} p_i^{(4,1)}},  & i\in \{5, 8, 11\},\\
\frac{m_i^{(4,2)}}{K^{(4,2)} p_i^{(4,2)}},  & i\in \{6, 9\};
\end{dcases}
\end{split}
\end{equation}
\textbf{Step 2}: For undirected orbit 3, we compute two estimates $\check{d}^{(3)}_v = \frac{m_3^{(4,1)}}{K^{(4,1)} p_3^{(4,1)}}$ and $\tilde{d}^{(3)}_v = \frac{m_3^{(3,2)}}{K^{(3,2)} p_3^{(3,2)}}$.
According to Theorem~\ref{theorem:estimatecardinality}, these are unbiased estimates of $d^{(3)}_v$
and their variances are
\begin{equation}\label{eq:varcheck_d_3}
\text{Var}(\check{d}^{(3)}_v) = \frac{d^{(3)}_v}{K^{(4,1)}}\left(\frac{1}{p_3^{(4,1)}} - d^{(3)}_v\right),
\end{equation}
\begin{equation}\label{eq:vartilde_d_3}
\text{Var}(\tilde{d}^{(3)}_v) = \frac{d^{(3)}_v}{K^{(3,2)}}\left(\frac{1}{p_3^{(3,2)}} - d^{(3)}_v\right).
\end{equation}
Theorem~\ref{theorem:mixestimators} allows us to compute the more accurate estimate
\begin{equation}\label{eq:undirectedhatd3mix}
\hat{d}^{(3)}_v = \lambda^{(3,1)}_v \check{d}^{(3)}_v + \lambda^{(3,2)}_v \tilde{d}^{(3)}_v,
\end{equation}
where $\lambda^{(3,1)}_v = \frac{\text{Var}(\tilde{d}^{(3)}_v)}{\text{Var}(\check{d}^{(3)}_v) + \text{Var}(\tilde{d}^{(3)}_v)}$
and $\lambda^{(3,2)}_v = \frac{\text{Var}(\check{d}^{(3)}_v)}{\text{Var}(\check{d}^{(3)}_v) + \text{Var}(\tilde{d}^{(3)}_v)}$
with $\text{Var}(\check{d}^{(3)}_v)$ and $\text{Var}(\tilde{d}^{(3)}_v)$
given by replacing $d^{(3)}_v$ with $\check{d}^{(3)}_v$ and $\tilde{d}^{(3)}_v$ in Eqs.~(\ref{eq:varcheck_d_3}) and~(\ref{eq:vartilde_d_3});

\noindent \textbf{Step 3}: For undirected orbit $i\in \{10, 12, 13, 14\}$, we use Theorem~\ref{theorem:estimatecardinality}
to compute two estimates $\check{d}^{(i)}_v = \frac{m_i^{(4,1)}}{K^{(4,1)} p_i^{(4,1)}}$ and $\tilde{d}^{(i)}_v = \frac{m_i^{(4,2)}}{K^{(4,2)} p_i^{(4,2)}}$
with variances
\begin{equation}\label{eq:varcheck_d_10}
\text{Var}(\check{d}^{(i)}_v) = \frac{d^{(i)}_v}{K^{(4,1)}}\left(\frac{1}{p_i^{(4,1)}} - d^{(i)}_v\right),
\end{equation}
\begin{equation}\label{eq:vartilde_d_10}
\text{Var}(\tilde{d}^{(i)}_v) = \frac{d^{(i)}_v}{K^{(4,2)}}\left(\frac{1}{p_i^{(4,2)}} - d^{(i)}_v\right).
\end{equation}
We then apply Theorem~\ref{theorem:mixestimators} to compute the more accurate estimate
\begin{equation}\label{eq:undirectedhatdmix}
\hat{d}^{(i)}_v = \lambda^{(i,1)}_v \check{d}^{(i)}_v + \lambda^{(i,2)}_v \tilde{d}^{(i)}_v,
\end{equation}
where $\lambda^{(i,1)}_v = \frac{\text{Var}(\tilde{d}^{(i)}_v)}{\text{Var}(\check{d}^{(i)}_v) + \text{Var}(\tilde{d}^{(i)}_v)}$
and $\lambda^{(i,2)}_v = \frac{\text{Var}(\check{d}^{(i)}_v)}{\text{Var}(\check{d}^{(i)}_v) + \text{Var}(\tilde{d}^{(i)}_v)}$
obtained by replacing $d^{(i)}_v$ with $\check{d}^{(i)}_v$ and $\tilde{d}^{(i)}_v$ in Eqs.~(\ref{eq:varcheck_d_10}) and~(\ref{eq:vartilde_d_10});

\noindent \textbf{Step 4}: For undirected orbit $i\in \{2, 4, 7\}$, we now estimate $d^{(i)}_v$ as
\[
\hat d^{(2)}_v  = \phi_v - \hat d^{(3)}_v,
\]
\begin{equation*}
\begin{split}
\hat d^{(4)}_v=&\Phi_v^{(3)} - 2 \hat d^{(3)}_v - 2 \hat d^{(8)}_v - 2 \hat d^{(9)}_v - \hat d^{(10)}_v - 4 \hat d^{(12)}_v - 2 \hat d^{(13)}_v\\
&- 6 \hat d^{(14)}_v,
\end{split}
\end{equation*}
\begin{equation*}
\hat d^{(7)}_v = \Phi_v^{(4)} - \hat d^{(11)}_v - \hat d^{(13)}_v - \hat d^{(14)}_v.
\end{equation*}

The following theorem presents the errors of the above estimates $\hat d^{(1)}_v, \ldots, \hat d^{(14)}_v$ for any $v$ in undirected graph $G$.
\begin{theorem}\label{theorem:error_undirected_GDV}
For $i\in \{1, \ldots, 14\}$, $\hat d^{(i)}_v$ is an unbiased estimate of $d^{(i)}_v$ with the following variance.\\
\textbf{(I)} For undirected orbit $i\in \{1, \ldots, 14\}\setminus \{2, 4, 7\}$, the variance of $\hat d^{(i)}_v$ is computed as 
\begin{equation*}
\begin{split}
\text{Var}(\hat{d}^{(i)}_v) &= 
\begin{dcases}
\frac{d^{(1)}_v}{K^{(3,2)}}\left(\frac{1}{p_1^{(3,2)}} - d^{(1)}_v\right),  & i=1,\\
\frac{d^{(i)}_v}{K^{(4,1)}}\left(\frac{1}{p_i^{(4,1)}} - d^{(i)}_v\right), & i\in \{5, 8, 11\},\\
\frac{d^{(i)}_v}{K^{(4,2)}}\left(\frac{1}{p_i^{(4,2)}} - d^{(i)}_v\right), & i\in \{6, 9\},\\
\frac{\text{Var}(\tilde{d}^{(i)}_v) \text{Var}(\check{d}^{(i)}_v)}{\text{Var}(\check{d}^{(i)}_v) + \text{Var}(\tilde{d}^{(i)}_v)}, & i\in \{3, 10, 12, 13, 14\},
\end{dcases}
\end{split}
\end{equation*}
where $\text{Var}(\tilde{d}^{(i)}_v)$ and $\text{Var}(\check{d}^{(i)}_v)$
are defined in Eqs.~(\ref{eq:varcheck_d_3}),~(\ref{eq:vartilde_d_3}),~(\ref{eq:varcheck_d_10}) and~(\ref{eq:vartilde_d_10}).\\
\textbf{(II))} For undirected orbit 2, the formula of $\text{Var}(\hat{d}^{(2)}_v)$ equals that of $\text{Var}(\hat{d}^{(3)}_v)$ derived above.\\ 
\textbf{(III)} For undirected orbit 4, $\text{Var}(\hat{d}^{(4)}_v)$ is computed as
\begin{equation}\label{eq:varhatd4}
\begin{split}
\text{Var}(\hat{d}^{(4)}_v) &= \sum_{j\in \{3, 8, 9, 10, 12, 13, 14\}} \chi_j^2 \text{Var}(\hat{d}^{(j)}_v)\\
&+\sum_{j, k\in \{3, 8, 9, 10, 12, 13, 14\}\wedge j\ne l} \chi_j \chi_l \text{Cov}(\hat{d}^{(j)}_v, \hat{d}^{(l)}_v),
\end{split}
\end{equation}
where $\chi_3 = \chi_8 = \chi_9 = \chi_{13} = 2$, $\chi_{10} = 1$, $\chi_{12} = 4$, and $\chi_{14} = 6$.\\
\textbf{(IV)} For undirected orbit 7, $\text{Var}(\hat{d}^{(7)}_v)$ is computed as
\begin{equation}\label{eq:varhatd7}
\begin{split}
\text{Var}(\hat{d}^{(7)}_v) &= \text{Var}(\hat{d}^{(11)}_v) + \text{Var}(\hat{d}^{(13)}_v) + \text{Var}(\hat{d}^{(14)}_v)\\
&+ \sum_{j, l\in\{11,13,14\}\wedge j\ne l} \text{Cov}(\hat{d}^{(j)}_v, \hat{d}^{(l)}_v).
\end{split}
\end{equation}
The covariances in the formulas of $\text{Var}(\hat{d}^{(4)}_v)$ and $\text{Var}(\hat{d}^{(7)}_v)$
(i.e., Eqs.~(\ref{eq:varhatd4}) and~(\ref{eq:varhatd7})) are computed as:\\
1. When $j, l\in \{5, 8, 11\}$ and $j\ne l$, $\text{Cov}(\hat{d}^{(j)}_v, \hat{d}^{(l)}_v) = -\frac{d^{(j)}_v d^{(l)}_v}{K^{(4,1)}}$;\\
2. When $j\in \{5, 8, 11\}$, $\text{Cov}(\hat{d}^{(j)}_v, \hat{d}^{(3)}_v) = \text{Cov}(\hat{d}^{(3)}_v, \hat{d}^{(j)}_v) =  -\frac{\lambda^{(3,1)}_v d^{(3)}_v d^{(j)}_v}{K^{(4,1)}}$;\\
3. When $j, l\in \{6, 9\}$ and $j\ne l$, $\text{Cov}(\hat{d}^{(j)}_v, \hat{d}^{(l)}_v) = -\frac{d^{(j)}_v d^{(l)}_v}{K^{(4,2)}}$;\\
4. When $j, l\in \{10, 12, 13, 14\}$ and $j\ne l$, we have $\text{Cov}(\hat{d}^{(j)}_v, \hat{d}^{(l)}_v) = -\frac{\lambda^{(j,1)}_v \lambda^{(l,1)}_v d^{(j)}_v d^{(l)}_v}{K^{(4,1)}}-\frac{\lambda^{(j,2)}_v \lambda^{(l,2)}_v d^{(j)}_v d^{(l)}_v}{K^{(4,2)}}$;\\
5. When $j\in \{3, 5, 8, 11\}$ and $l\in \{6, 9\}$, we have $\text{Cov}(\hat{d}^{(j)}_v, \hat{d}^{(l)}_v) = \text{Cov}(\hat{d}^{(l)}_v, \hat{d}^{(j)}_v) =0$;\\
6. When $j \in \{5, 8, 11\}$ and $l\in \{10, 12, 13, 14\}$, $\text{Cov}(\hat{d}^{(j)}_v, \hat{d}^{(l)}_v) = \text{Cov}(\hat{d}^{(l)}_v, \hat{d}^{(j)}_v) = -\frac{\lambda^{(l,1)}_v d^{(j)}_v d^{(l)}_v}{K^{(4,1)}}$;\\
7. When $j \in \{6, 9\}$ and $l\in \{10, 12, 13, 14\}$, $\text{Cov}(\hat{d}^{(j)}_v, \hat{d}^{(l)}_v) = \text{Cov}(\hat{d}^{(l)}_v, \hat{d}^{(j)}_v) = -\frac{\lambda^{(l,2)}_v d^{(j)}_v d^{(l)}_v}{K^{(4,2)}}$;\\
8. When $j \in \{10, 12, 13, 14\}$,
$\text{Cov}(\hat{d}^{(3)}_v, \hat{d}^{(j)}_v) = \text{Cov}(\hat{d}^{(j)}_v, \hat{d}^{(3)}_v) = -\frac{\lambda^{(3,1)}_v \lambda^{(j,1)}_v d^{(3)}_v d^{(j)}_v}{K^{(4,1)}}$.
\end{theorem}

\section{SAND-3D: Estimation of Directed Orbit Degrees}\label{sec:directedgraphestimation}
Due to a large number of directed 4-node graphlets and orbits,
in this paper we focus on 3-node directed graphlets and orbits.
Next, we introduce our method for estimating 3-node directed orbit degrees.

\subsection{Estimating Single Directed Orbit Degree}
For a directed orbit $i$,
denote $unorbit(i)$ as its associated undirected orbit when discarding the directions of edges in the graphlet.
For example, directed orbits 2, 4, 5, 7, 9, 10, 12, 13, and 15 in Fig.~\ref{fig:othergraphlets} are associated with undirected orbit 1 in Fig.~\ref{fig:GDDgraphlets},
directed orbits 1, 3, 6, 8, 11, and 14 in Fig.~\ref{fig:othergraphlets} are associated with undirected orbit 2 in Fig.~\ref{fig:GDDgraphlets},
and directed orbits 16--30 in Fig.~\ref{fig:othergraphlets} are associated with undirected orbit 3 in Fig.~\ref{fig:GDDgraphlets}.
Given a sampling method from Section~\ref{sec:preliminaries},
the probability of it sampling a CIS in directed orbit $i$, denoted  by $p_i$,
equals the probability of the method sampling undirected orbit $unorbit(i)$ derived in Section~\ref{sec:preliminaries}.
When undirected orbit $unorbit(i)$ can only sampled by one method in Section~\ref{sec:preliminaries}
(e.g., Randgraf-3-2 is the only one that can sample $unorbit(i) = 1$),
we use the method to obtain $K$ CISes that include a node $v\in V$.
Let $m_i$ denote the number of sampled CISes that include $v$ in directed orbit $i$.
According to Theorem~\ref{theorem:estimatecardinality},
we estimate $d^{(i, \text{dir})}_v $ as
\begin{equation*}
\hat d^{(i, \text{dir})}_v = \frac{m_i}{K p_i}
\end{equation*}
with variance $\text{Var}(\hat d^{(i, \text{dir})}_v) = \frac{d^{(i, \text{dir})}_v}{K}\left(\frac{1}{p_i} - d^{(i, \text{dir})}_v\right)$.
When more than one method is able to sample undirected orbit $unorbit(i)$,
we select the most efficient method, the one with the smallest $\frac{\text{Var}(\hat d^{(i, \text{dir})}_v)}{K t_v}$ to estimate $d^{(i, \text{dir})}_v$,
where $t_v$ is the average computational time of the method sampling a CIS, which is shown in Table~\ref{tab:samplingmethods}.

\subsection{Estimating all Directed Orbit Degrees}
We develop method SAND-3D consisting of both Randgraf-3-1 and Randgraf-3-2 to estimate all 3-node directed orbit degrees $d^{(1,\text{dir})}_v, \ldots, d^{(30,\text{dir})}_v$.
Directed orbit $i\in\{1, 3, 6, 8, 11, 14\}$  can be sampled by Randgraf-3-1 but not Randgraf-3-2, so we compute $d^{(i, \text{dir})}_v$ as the unbiased estimate given by Randgraf-3-1.
Directed orbit $i\in\{2, 4, 5, 7, 9, 10, 12, 13, 15\}$ can be sampled by Randgraf-3-2 but not Randgraf-3-1,
so we compute $d^{(i, \text{dir})}_v$ as the unbiased estimate given by Randgraf-3-2.
We estimate $d^{(i, \text{dir})}_v$ for directed orbit $i\in \{16, 17, \ldots, 30\}$, by combining two unbiased estimates given by Randgraf-3-1 and Randgraf-3-2
according to Theorem~\ref{theorem:mixestimators}.

\section{Evaluation} \label{sec:results}

\begin{figure*}[htb]
\center
\subfigure[real values\label{fig:groundtruthundirected}]{
\includegraphics[width=0.49\textwidth]{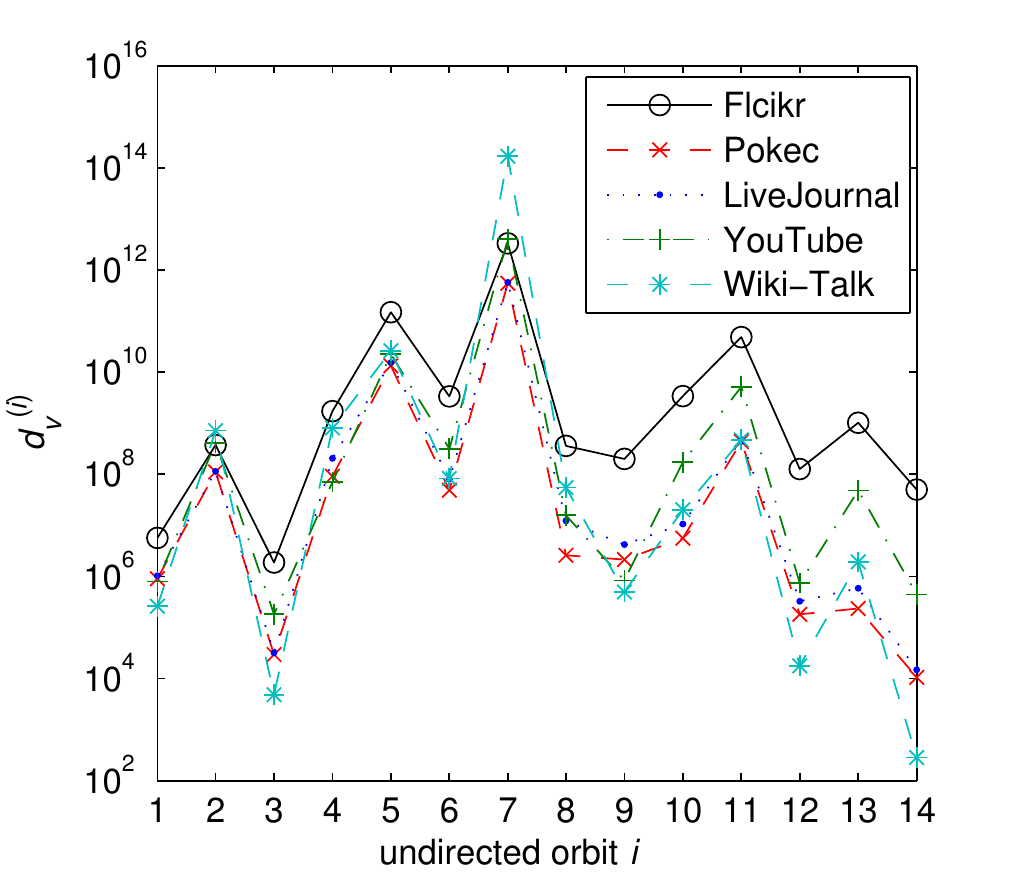}}
\subfigure[NRMSEs of our method SAND\label{fig:nrmseundirected}]{
\includegraphics[width=0.49\textwidth]{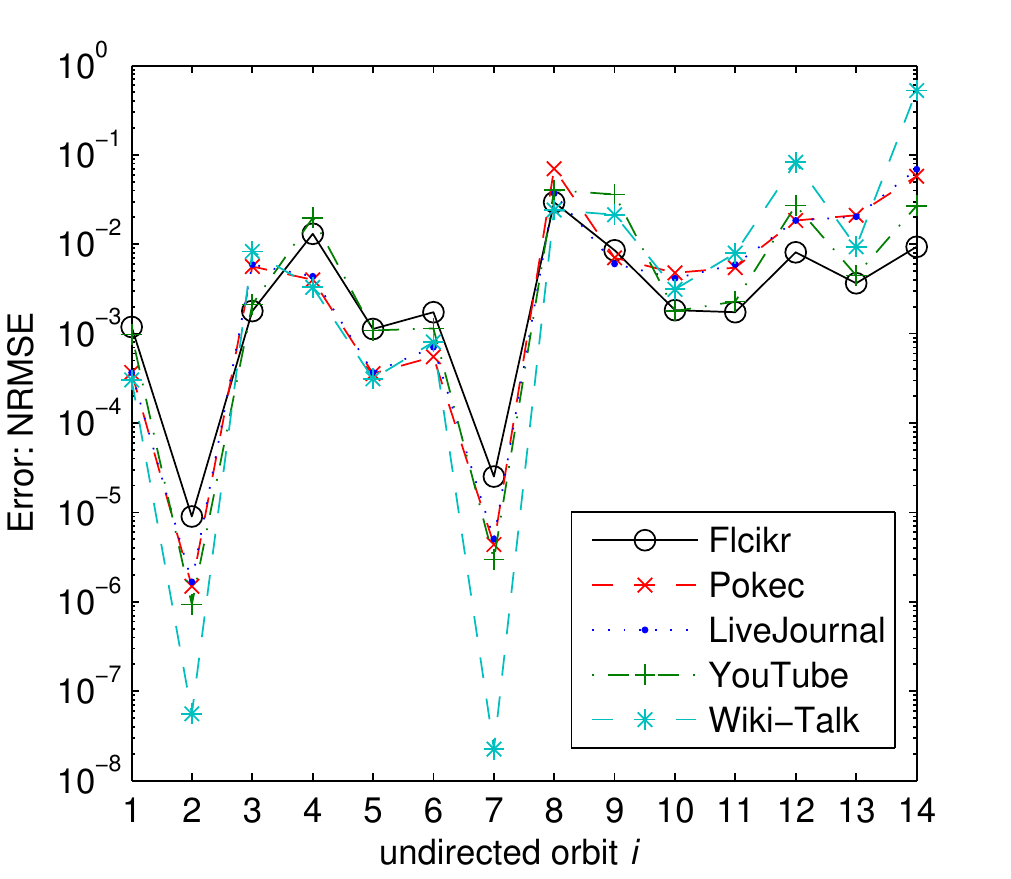}}
\caption{Real values and NRMSEs of our estimates of 3- and 4-node undirected orbit degrees of node $v_\text{max}$.}
\label{fig:resultundirected}
\end{figure*}

\begin{figure*}[htb]
\center
\includegraphics[width=\textwidth]{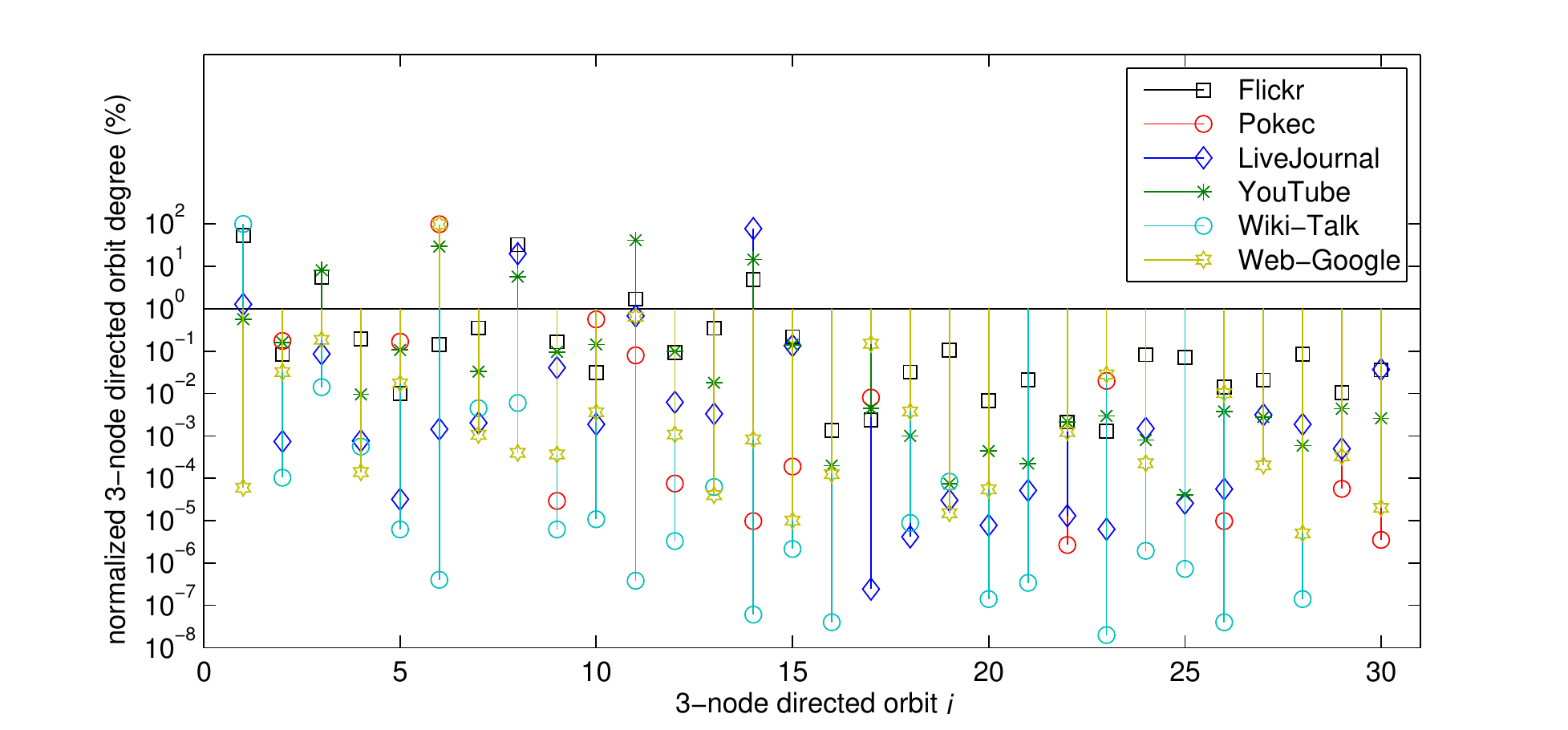}
\caption{Real values of normalized 3-node directed orbit degrees of node $v_\text{max}$,
i.e., $\frac{d_{v_\text{max}}^{(i,\text{dir})}}{\sum_{j=1}^{30} \hat d_{v_\text{max}}^{(j,\text{dir})}}\times 100\%$,
$1\le i\le 30$.}
\label{fig:ground3directedtruth}
\end{figure*}

\begin{figure*}[htb]
\center
\subfigure[Flickr. Top-10 directed orbits with the largest orbit degrees: 1, 8, 3, 14, 11, 7, 13, 15, 4, and 9.\label{fig:3flickr}]{
\includegraphics[width=0.325\textwidth]{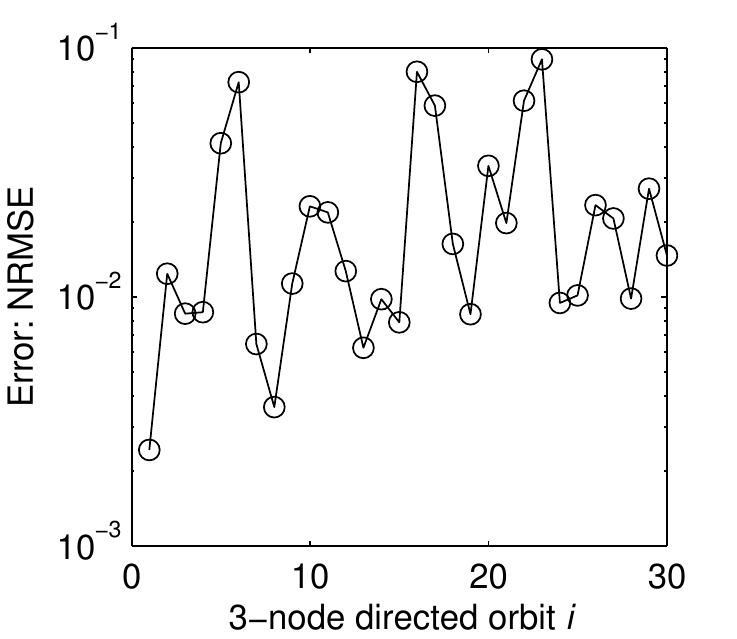}}
\subfigure[Pokec. Top-10 directed orbits with the largest orbit degrees: 11, 6, 14, 3, 8, 1, 2, 10, 15, and 5.\label{fig:3pokec}]{
\includegraphics[width=0.325\textwidth]{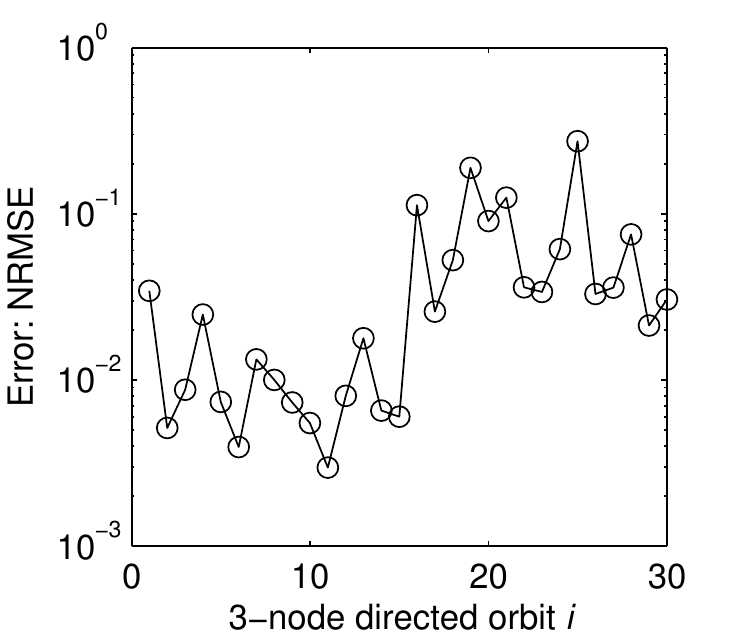}}
\subfigure[YouTube. Top-10 directed orbits with the largest orbit degrees: 14, 8, 1, 11, 15, 3, 9, 30, 12, and 13.\label{fig:3youtube}]{
\includegraphics[width=0.325\textwidth]{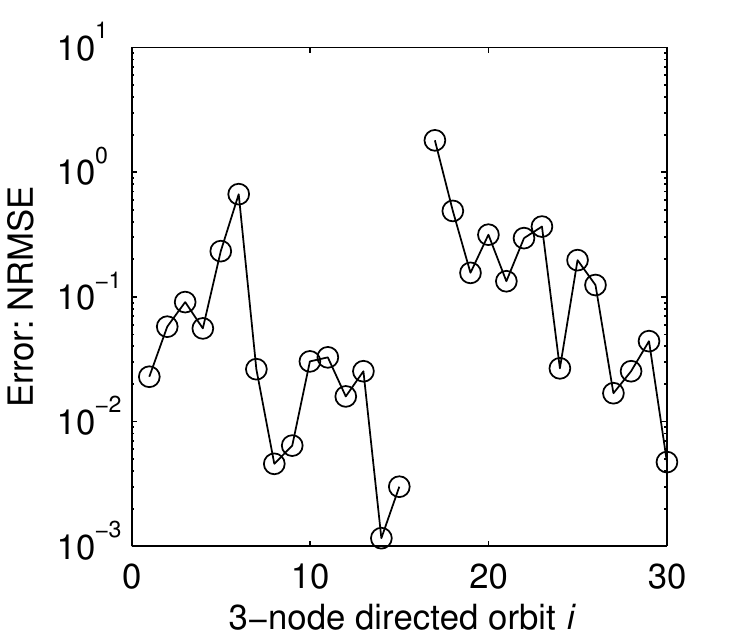}}
\subfigure[LiveJournal. Top-10 directed orbits with the largest orbit degrees: 6, 10, 2, 5, 11, 23, 17, 15, 12, and 29.\label{fig:3livejournal}]{
\includegraphics[width=0.325\textwidth]{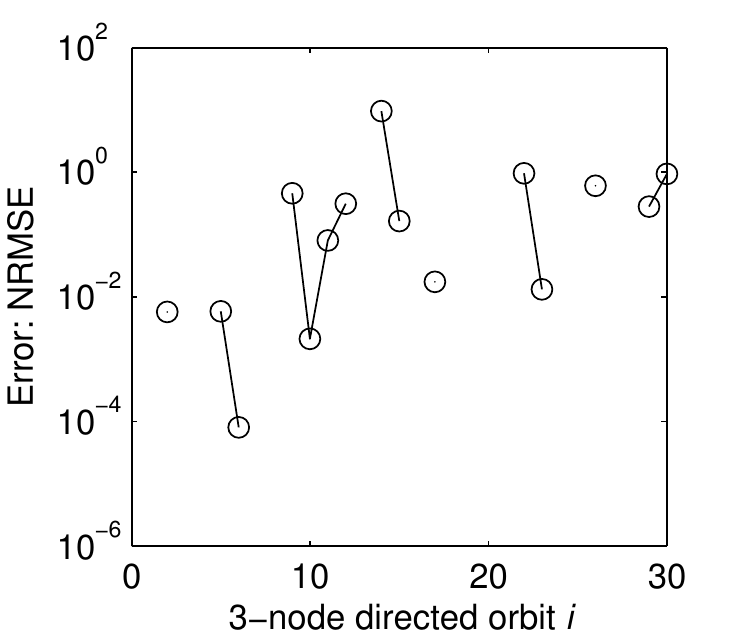}}
\subfigure[Wiki-Talk. Top-10 directed orbits with the largest orbit degrees: 1, 3, 8, 7, 4, 2, 19, 13, 10, and 18.\label{fig:3wikitalk}]{
\includegraphics[width=0.325\textwidth]{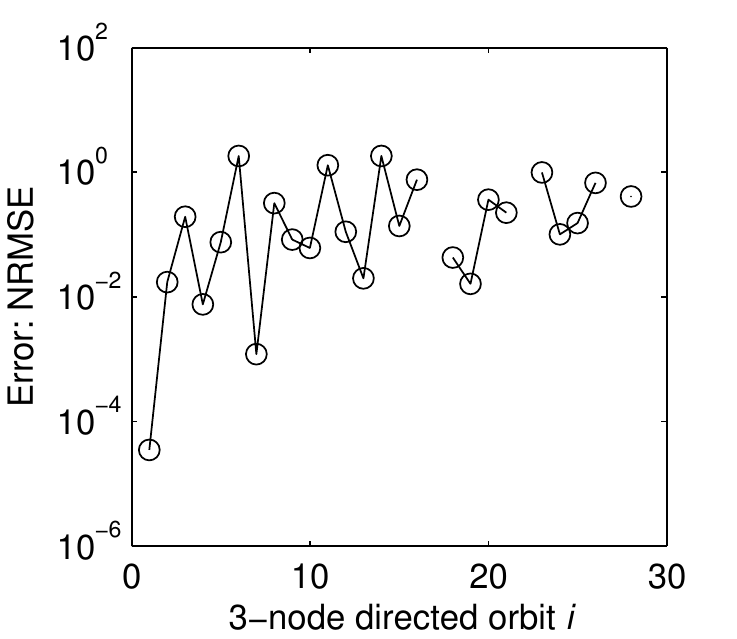}}
\subfigure[Web-Google. Top-10 directed orbits with the largest orbit degrees: 6, 11, 3, 17, 2, 23, 5, 26, 18, and 10.\label{fig:3webgoogle}]{
\includegraphics[width=0.325\textwidth]{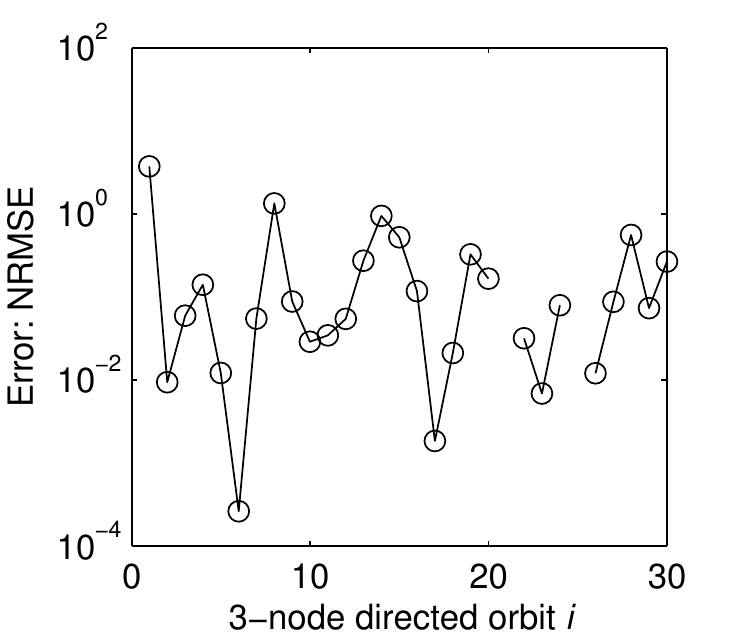}}
\caption{NRMSEs of our estimates of 3-node directed orbit degrees.}
\label{fig:result3directed}
\end{figure*}

\subsection{Datasets}
We perform our experiments on the following publicly available datasets taken from the Stanford Network Analysis Platform (SNAP)\footnote{www.snap.stanford.edu},
which are summarized in Table~\ref{tab:datasets}.
We evaluate our method for computing the orbit degrees of node $v_{\text{max}}$ with the largest degree in the graph of interest.
\begin{table}[htb]
\begin{center}
\caption{Graph datasets used in our experiments. "edges" refers to the number of edges in the
undirected graph generated by discarding edge directions. "max-degree" represents  the maximum number of edges incident to a node in the undirected graph.\label{tab:datasets}}
\begin{tabular}{|c|ccc|}
\hline
graph&nodes&edges&max-degree\\
\hline
Flickr~\cite{MisloveIMC2007}&1,715,255&15,555,041&27,236\\
Pokec~\cite{Takac2012}&1,632,803&22,301,964&14,854\\
LiveJournal~\cite{MisloveIMC2007}&5,189,809&48,688,097&15,017\\
YouTube~\cite{MisloveIMC2007}&1,138,499&2,990,443&28,754\\
Wiki-Talk~\cite{Leskovec2010}&2,394,385&4,659,565&100,029\\
Web-Google~\cite{GoogleProgrammingContest2002}&875,713&	4,322,051&6,332\\
%soc-Epinions1~\cite{Richardson2003}&75,897&405,740&3,044\\
%soc-Slashdot08~\cite{LeskovecIM2009}&77,360&469,180&2,539\\
%com-DBLP~\cite{YangICDM2012}&317,080&1,049,866&343\\
%com-Amazon~\cite{YangICDM2012}&334,863&925,872&549\\
%p2p-Gnutella08~\cite{Ripeanu2002}&6,301&20,777&97\\
%ca-GrQc~\cite{LeskovecTKDD2007}&5,241&14,484&81\\
%ca-CondMat~\cite{LeskovecTKDD2007}&23,133&93,439&279\\
%ca-HepTh~\cite{LeskovecTKDD2007}&9,875&25,937&65\\
\hline
\end{tabular}
\end{center}
\end{table}

\subsection{Metric}
We use the normalized root mean square error (NRMSE) to measure the relative error of
the orbit degree estimate $\hat d_{v_{\text{max}}}^{(i)}$ with respect to its true value $d_{v_{\text{max}}}^{(i)}$, $i=1,2,\dots$.
It is defined as:
\[
\text{NRMSE}(\hat d_{v_{\text{max}}}^{(i)})=\frac{\sqrt{\text{MSE}(\hat d_{v_{\text{max}}}^{(i)})}}{d_{v_{\text{max}}}^{(i)}}, \qquad i=1, 2, \dots,
\]
where $\text{MSE}(\hat d_{v_{\text{max}}}^{(i)})$ denotes the mean square error of $\hat d_{v_{\text{max}}}^{(i)}$:
\begin{equation*}
\begin{split}
\text{MSE}(\hat d_{v_{\text{max}}}^{(i)})&=\mathbb{E}((\hat d_{v_{\text{max}}}^{(i)}- d_{v_{\text{max}}}^{(i)})^2)\\
&=\text{Var}(\hat d_{v_{\text{max}}}^{(i)})+\left(\mathbb{E}(\hat d_{v_{\text{max}}}^{(i)})-d_{v_{\text{max}}}^{(i)}\right)^2.
\end{split}
\end{equation*}
$\text{MSE}(\hat d_{v_{\text{max}}}^{(i)})$ decomposes into a sum of the variance and bias of the estimator $\hat d_{v_{\text{max}}}^{(i)}$, both quantities are important and need to be as small as possible to achieve good estimation performance.
When $\hat d_{v_{\text{max}}}^{(i)}$ is an unbiased estimator of $d_{v_{\text{max}}}^{(i)}$,
we have $\text{MSE}(\hat d_{v_{\text{max}}}^{(i)})= \text{Var}(\hat d_{v_{\text{max}}}^{(i)})$.
In our experiments, we average the estimates and calculate their NRMSEs over 1,000 runs.
We evenly distribute the sampling budget among the sampling methods of SAND and SAND-3D, and leave the optimal budget distribution in future study.
Our experiments are conducted on a server with a Quad-Core AMD Opeteron (tm) 8379 HE CPU 2.39 GHz processor and 128 GB DRAM memory.

\subsection{Results}
\subsubsection{Estimating undirected orbit degrees}
We evaluate the performance of SAND
by comparing its performance to the state-of-the-art enumeration method 4-Prof-Dist~\cite{ElenbergWWW2016}
for estimating 3- and 4-node undirected orbit degrees over the undirected graphs of datasets Flickr, Pokec, LiveJounal, YouTube, and Wiki-Talk,
which are obtained by discarding edge directions.
Table~\ref{tab:timeundirected} shows that with a sampling budget $10^6$ SAND is 183, 3.9, 15, and 81 times faster than 4-Prof-Dist
for computing 3- and 4-node undirected orbit degrees of graphs Flickr, Pokec, LiveJounal, YouTube, and Wiki-Talk respectively.
Fig.~\ref{fig:groundtruthundirected} shows the real values of 3- and 4-node undirected orbit degrees of $v_\text{max}$.
Roughly speaking, 3- and 4-node undirected orbit degree distributions of graphs Flickr, Pokec, LiveJounal, YouTube, and Wiki-Talk exhibit similar patterns.
$d_{v_{\text{max}}}^{(7)}$, $d_{v_{\text{max}}}^{(5)}$, and $d_{v_{\text{max}}}^{(11)}$ are the three largest 3- and 4-node undirected orbit degrees.
Fig.~\ref{fig:nrmseundirected} shows the NRMSEs of our estimates $\hat d_{v_{\text{max}}}^{(i)}$, $i=1, \ldots, 14$.
We observe that all NRMSEs of $\hat d_{v_{\text{max}}}^{(i)}$ are smaller than 0.1 except the NRMSE of $\hat d_{v_{\text{max}}}^{(14)}$.
The NRMSEs of Top-3 orbits degrees $\hat d_{v_{\text{max}}}^{(7)}$, $\hat d_{v_{\text{max}}}^{(5)}$, and $\hat d_{v_{\text{max}}}^{(11)}$ are smaller than 0.01.

\begin{table}[htb]
\begin{center}
\caption{Computational cost of computing 3- and 4-node undirected orbit degrees of node $v_\text{max}$.\label{tab:timeundirected}}
\begin{tabular}{|c|r|r|} \hline
\multirow{2}{*}{\textbf{graph}}&\multicolumn{2}{c|}{\textbf{computational time (seconds)}}\\
\cline{2-3}
&4-Prof-Dist~\cite{ElenbergWWW2016}&SAND\\\hline
Flickr&7,681&41.9\\ \hline
Pokec&179&45.7\\ \hline
LiveJournal&300&58.2\\ \hline
YouTube&675&45.3\\ \hline
Wiki-Talk&3,489&43.0\\ \hline
\end{tabular}
\end{center}
\end{table}

\subsubsection{Estimating directed orbit degrees}
To the best of our knowledge, there exist no sampling method for estimating 3-node directed orbit degrees.
Therefore, we evaluate the performance of SAND-3D in comparison with
the method of enumerating and classifying all the 3-node CISes that include $v_{\text{max}}$.
Table~\ref{tab:timedirected} shows that with a sampling budget $10^6$ SAND-3D is 229, 76.1, 246, 36,034 and 12.7 times faster than the enumeration method
for computing 3-node directed orbit degrees of graphs Flickr, Pokec, LiveJounal, YouTube, Wiki-Talk, and Web-Google respectively.
Let $\frac{d_{v_\text{max}}^{(i,\text{dir})}}{\sum_{j=1}^{30} \hat d_{v_\text{max}}^{(j,\text{dir})}}\times 100\%$
denote the normalized 3-node directed orbit $i$ degrees of $v_\text{max}$, $1\le i\le 30$.
Fig.~\ref{fig:ground3directedtruth} shows the real values of normalized 3-node directed orbit degrees of $v_\text{max}$.
We observe that the 3-node directed orbit degrees of $v_\text{max}$ exhibit quite different patterns for different graphs.
For example, Flickr and Wiki-Talk have the largest graphlet degree in directed orbit 1,
Pokec and Web-Google have the largest graphlet degree in directed orbit 6,
LiveJournal has the largest graphlet degree in directed orbit 14,
and YouTube has the largest graphlet degree in directed orbit 11.
Fig.~\ref{fig:result3directed} shows the NRMSEs of our estimates $\hat d_{v_{\text{max}}}^{(i,\text{dir})}$, $i=1, \ldots, 30$.
We observe that the NRMSEs of estimates of the ten largest orbit degrees are smaller than 0.1 for all the graphs studied in this paper.

Although the NRMSEs of small orbit degrees exhibit large errors,
we observe that SAND-3D is accurate enough for applications such as detecting
the most frequent orbits, i.e., the orbits with the largest orbit degrees.
Table~\ref{tab:topk} shows the results of detecting the five, ten, and fifteen most frequent directed orbits.
We can see that SAND-3D successfully identifies all the five and ten most frequent directed orbits.
On average, no more than one of the fifteen most frequent directed orbits is missed by SAND-3D.
We also study the $L_1$ and $L_2$ distances between our estimates and real values, which are defined as $L_1 =\sum_{i=1}^{30} |\frac{\hat d_{v_\text{max}}^{(i,\text{dir})}}{\sum_{j=1}^{30} \hat d_{v_\text{max}}^{(j,\text{dir})}}- \frac{d_{v_\text{max}}^{(i,\text{dir})}}{\sum_{j=1}^{30} d_{v_\text{max}}^{(j,\text{dir})}}|$
 and $L_2=\sum_{i=1}^{30} \sqrt{(\frac{\hat d_{v_\text{max}}^{(i,\text{dir})}}{\sum_{j=1}^{30} \hat d_{v_\text{max}}^{(j,\text{dir})}}- \frac{d_{v_\text{max}}^{(i,\text{dir})}}{\sum_{j=1}^{30} d_{v_\text{max}}^{(j,\text{dir})}})^2}$.
Table~\ref{tab:L2L1} shows that $L_1$ and $L_2$ distances are smaller than 0.001 and 0.002 respectively.
This indicates that estimates given by SAND-3D are accurate for $L_1$ and $L_2$ distances based machine learning applications.

\begin{table}[htb]
\begin{center}
\caption{Computational cost of computing 3-node directed orbit degrees of node $v_\text{max}$.\label{tab:timedirected}}
\begin{tabular}{|c|r|r|} \hline
\multirow{2}{*}{\textbf{graph}}&\multicolumn{2}{c|}{\textbf{computational time (seconds)}}\\
\cline{2-3}
&enumeration method&SAND-3D\\\hline
Flickr&1,461&6.38\\ \hline
Pokec&367&4.82\\ \hline
LiveJournal&472&6.69\\ \hline
YouTube&1,294&5.26\\ \hline
Wiki-Talk&181,609&5.04\\ \hline
Web-Google&61.7&4.87\\ \hline
\end{tabular}
\end{center}
\end{table}

\begin{table}[htb]
\begin{center}
\caption{Accuracy of identifying the five, ten, and fifteen most frequent 3-node directed orbits of $v_\text{max}$.\label{tab:topk}}
\begin{tabular}{|c|c|c|c|} \hline
\multirow{2}{*}{\textbf{graph}}&\multicolumn{3}{c|}{\textbf{\# Top frequent orbits correctly detected}}\\
\cline{2-4}
&Top-5&Top-10&Top-15\\\hline
Flickr&5&10&14.9\\ \hline
Pokec&5&10&15.0\\ \hline
LiveJournal&5&10&14.0\\ \hline
YouTube&5&10&14.5\\ \hline
Wiki-Talk&5&10&15.0\\ \hline
Web-Google&5&10&14.6\\ \hline
\end{tabular}
\end{center}
\end{table}

\begin{table}[htb]
\begin{center}
\caption{Errors $L_2=\sum_{i=1}^{30} \sqrt{(\frac{\hat d_{v_\text{max}}^{(i,\text{dir})}}{\sum_{j=1}^{30} \hat d_{v_\text{max}}^{(j,\text{dir})}}- \frac{d_{v_\text{max}}^{(i,\text{dir})}}{\sum_{j=1}^{30} d_{v_\text{max}}^{(j,\text{dir})}})^2}$
and $L_1 =\sum_{i=1}^{30} |\frac{\hat d_{v_\text{max}}^{(i,\text{dir})}}{\sum_{j=1}^{30} \hat d_{v_\text{max}}^{(j,\text{dir})}}- \frac{d_{v_\text{max}}^{(i,\text{dir})}}{\sum_{j=1}^{30} d_{v_\text{max}}^{(j,\text{dir})}}|$.
\label{tab:L2L1}}
\begin{tabular}{|c|r|r|r|r|} \hline
\multirow{2}{*}{\textbf{graph}}&\multicolumn{2}{c|}{$L_2$}&\multicolumn{2}{c|}{$L_1$}\\
\cline{2-5}
&mean&variance&mean&variance\\\hline
Flickr&9.1e-04&2.5e-07&1.8e-03&7.1e-07\\ \hline
Pokec&1.1e-03&1.8e-07&2.1e-03&5.9e-07\\ \hline
LiveJournal&4.1e-05&9.0e-10&7.0e-05&2.1e-09\\ \hline
YouTube&6.2e-04&1.5e-07&1.1e-03&3.5e-07\\ \hline
Wiki-Talk&2.3e-05&1.7e-10&3.6e-05&4.1e-10\\ \hline
Web-Google&1.6e-04&9.6e-09&2.6e-04&2.3e-08\\ \hline
\end{tabular}
\end{center}
\end{table}

\section{Related Work} \label{sec:related}
Recently, a number of efforts have focused on designing sampling methods for computing a large graph's graphlet concentrations~\cite{Alon1995,Kashtan2004,Wernicke2006,OmidiGenes2009,Bhuiyan2012,GRAFT2012,TKDDWang2014}
and graphlet counts~\cite{Alon1995,TsourakakisKDD2009,PavanyVLDB2013,JhaKDD2013,AhmedKDD2014,JhaWWW2015}.
To estimate graphlet concentrations, Kashtan et al.~\cite{Kashtan2004} proposed a simple subgraph sampling method.
However their method is computationally expensive when calculating the weight of each sampled subgraph, which is used for correcting bias introduced by edge sampling.
To address this drawback, Wernicke~\cite{Wernicke2006} proposed a method named FANMOD based on enumerating subgraph trees.
GUISE proposed a Metropolis-Hastings based sampling method to estimate 3-node, 4-node, and 5-node graphlet concentrations\footnote{The concentration of a particular $k$-node graphlet in a network refers to the ratio of the graphlet count to the total number of $k$-node CISes in the network, $k=3, 4, 5, \ldots$.}.
These methods assume the entire topology of the graph of interest is known in advance and it can be fit into the memory.
Wang et al.~\cite{TKDDWang2014} propose an efficient crawling method to estimate online social network motif concentrations,
when the graph's topology is not available in advance and it is costly to sample the entire topology.
When the available dataset is a set of random edges sampled from streaming graphs\footnote{Streaming graph is given in form of a stream of edges.}, Wang et al.~\cite{PinghuiMotifEdgeSampling2014} propose an efficient crawling method to estimate graphlet concentrations.

The above methods fail to compute graphlet counts, which is more fundamental than graphlet concentrations.
Alon et al.~\cite{Alon1995} propose a color-coding method to reduce the computational cost of counting subgraphs.
Color-coding reduces computation by coloring nodes randomly and enumerating only colorful CISes (i.e., CISes that consist of nodes with distinct colors),
but~\cite{JhaWWW2015} reveals that the color-coding method is not scalable and is hindered by the sheer number of colorful CISes.
\cite{TsourakakisKDD2009,PavanyVLDB2013,JhaKDD2013,AhmedKDD2014} develop sampling methods to estimate the number of triangles of static and dynamic graphs.
Jha et al.~\cite{JhaWWW2015} develop sampling methods to estimate counts of 4-node undirected graphlets.
Wang et al.~\cite{WangMOSSTech15} develop a sampling method to estimate counts of 5-node undirected motifs.
These methods are designed to sample all subgraphs,
but not tailored to meet the need of sampling the subgraphs \textbf{that include a given node}.
Elenberg et al.~\cite{ElenbergWWW2016} develop a method to estimate counts of 4-node undirected motifs that include a given node based on random edge sampling,
but their sampling method cannot be used to estimate node orbit degrees because the orbit of a node in a sampled CIS may be different from that of the node in the original CISes.
We point out that method 4-Prof-Dist in~\cite{ElenbergWWW2016} can be easily extended and used to compute the exact values of a node's 4-node undirected orbit degrees, but it fails to compute directed orbit degrees.
To the best of our knowledge, we are the first to propose sampling methods for estimating a node's orbit degrees for large graphs.

\section{Conclusions and Future Work} \label{sec:conclusions}
We develop computationally efficient sampling methods to estimate the counts of 3- and 4-node undirected and directed graphlet orbit degrees for large graphs.
We provide unbiased estimators of graphlet orbit degrees, and derive simple and exact formulas for the variances of the estimators.
Meanwhile, we conduct experiments on a variety of publicly available datasets,
and experimental results show that our methods accurately estimates graphlet orbit degrees for the nodes with the largest degrees in graphs with millions of edges within one minute.
In future, we plan to extend SAND to estimate 5-node (or even higher order) graphlet orbit degrees and investigate the graphlet orbit degree signatures as features for various learning tasks.

\section*{Acknowledgment}
This work was supported in part by ARL under Cooperative Agreement W911NF-09-2-0053. The views and conclusions contained in this document are those of the authors and should not be interpreted as representing the official policies, either expressed or implied of the ARL, or the U.S. Government.
This work was supported in part by the National Natural Science Foundation of China (61103240, 61103241, 61221063, 61221063, 91118005, U1301254),
the 111 International Collaboration Program of China, 863 High Tech Development Plan (2012AA011003), the Prospective Research Project
on Future Networks of Jiangsu Future Networks Innovation Institute, and
the Application Foundation Research Program of SuZhou (SYG201311).
The work of John C.S. Lui was supported in part by the GRF 415013.

\section*{Appendix}
\subsection*{Implementation Details}\label{subsection:implementation}
We discuss our methods for implementing the functions in the Algorithms
we presented subsections~\ref{sec: 3-node-sampling} and~\ref{sec: 4-node-sampling}.
We also analyze their computational complexities.

\textbf{Initialization of $\phi_v$, $\varphi_v$, $\Phi_v^{(1)}$, $\Phi_v^{(2)}$, $\Phi_v^{(3)}$, and $\Phi_v^{(4)}$}:
For each node $v$, we store its degree $d_v$ and store its neighbors' degrees in a list.
Therefore, $O(1)$ and $O(d_v)$ operations are required to compute $\phi_v$ and $\varphi_v$ respectively.
Similarly, one can easily find that $O(N_v)$, $O(N_v)$, $O(\sum_{u\in N_v} d_u)$, and $O(1)$ operations are required to compute $\Phi_v^{(1)}$,
$\Phi_v^{(2)}$, $\Phi_v^{(3)}$, and $\Phi_v^{(4)}$ respectively.

\textbf{$\text{RandomVertex}(N_v)$}: We use an array $N_v[1, \ldots, d_v]$ to store the neighbors of $v$.
Function $\text{RandomVertex}(N_v\setminus \{u\})$ first randomly selects number $rnd$ from $\{1, \ldots, d_v\}$ and then returns node $N_v[rnd]$.
Its computational complexity is just $O(1)$.

\textbf{$\text{RandomVertex}(N_v\setminus \{u\})$}: Let $POS_{v,u}$ denote the index of $u$ in the list $N_v[1, \ldots, d_v]$, i.e., $N_v[POS_{v,u}]=u$.
Then, function $\text{RandomVertex}(N_v\setminus \{u\})$ includes the following steps:
\begin{itemize}
  \item \textbf{Step 1}: Select number $rnd$ from $\{1, \ldots, d_v\}\setminus\{POS_{v,u}\}$ at random;
  \item \textbf{Step 2}: Return $N_v[rnd]$.
\end{itemize}
Its computational complexity is $O(1)$.

%\textbf{$\text{RandomVertex}(N_u-\{v\})$ and $\text{RandomVertex}(N_w-\{u\})$}:  They are achieved similarly to that for $\text{RandomVertex}(N_v-\{u\})$.

\textbf{$\text{RandomVertex}(N_v\setminus \{u, w\})$}: Similarly, $\text{RandomVertex}(N_v\setminus \{u, w\})$ includes the following steps:
\begin{itemize}
  \item \textbf{Step 1}: Select number $rnd$ from\\
  $\{1, \ldots, d_v\}\setminus \{POS_{v,u}, POS_{v,w}\}$ at random;
  \item \textbf{Step 2}: Return $N_v[rnd]$.
\end{itemize}
Its computational complexity is $O(1)$.

\textbf{$\text{WeightRandomVertex}(N_v, \alpha^{(v)})$}:
We store an array $ACC\_\alpha^{(v)}$ in memory,
where $ACC\_\alpha^{(v)}[i]$ is defined as $ACC\_\alpha^{(v)}[i] = \sum_{j=1}^i (d_{N_v[j]} - 1)$, $1\le i\le d_v$.
Let $ACC\_\alpha^{(v)}[0]=0$.
Then, $\text{WeightRandomVertex}(N_v, \alpha^{(v)})$ includes the following steps:
\begin{itemize}
  \item \textbf{Step 1}: Select number $rnd$ from $\{1, \ldots, ACC\_\alpha^{(v)}[d_v]\}$ at random;
  \item \textbf{Step 2}: Find $i$ such that
  \[
  ACC\_\alpha^{(v)}[i-1] < rnd \le ACC\_\alpha^{(v)}[i],
  \]
which is solved by binary search;
  \item \textbf{Step 3}: Return $N_v[i]$.
\end{itemize}
Its computational complexity is $O(\log d_v)$.

\textbf{$\text{WeightRandomVertex}(N_v, \beta^{(v)})$}: We store an array $ACC\_\beta^{(v)}$ in memory,
where $ACC\_\beta^{(v)}[i]$ is defined as $ACC\_\beta^{(v)}[i] = \sum_{j=1}^i (\phi_{N_v[j]}  - d_{N_v[j]} + 1)$, $1\le i\le d_v$.
Let $ACC\_\beta^{(v)}[0]=0$.
Then, $\text{WeightRandomVertex}(N_v, \beta^{(v)})$ includes the following steps:
\begin{itemize}
  \item \textbf{Step 1}: Select number $rnd$ from $\{1, \ldots, ACC\_\beta^{(v)}[d_v]\}$ at random;
  \item \textbf{Step 2}: Find $i$ such that
  \[
  ACC\_\beta^{(v)}[i-1] < rnd \le ACC\_\beta^{(v)}[i],
  \]
which again is solved by binary search;
  \item \textbf{Step 3}: Return $N_v[i]$.
\end{itemize}
Its computational complexity is $O(\log d_v)$.

\textbf{$\text{WeightRandomVertex}(N_v, \gamma^{(v)})$}:  We store an array $ACC\_\gamma^{(v)}$ in memory,
where $ACC\_\gamma^{(v)}[i]$ is defined as $ACC\_\gamma^{(v)}[i] = \sum_{j=1}^i (\varphi_{N_v[j]}  - d_v + 1)$, $1\le i\le d_v$.
Let $ACC\_\gamma^{(v)}[0]=0$.
Then, $\text{WeightRandomVertex}(N_v, \gamma^{(v)})$ includes the following steps:
\begin{itemize}
  \item \textbf{Step 1}: Select number $rnd$ from $\{1, \ldots, ACC\_\gamma^{(v)}[d_v]\}$ at random;
  \item \textbf{Step 2}: Find $i$ such that
  \[
  ACC\_\gamma^{(v)}[i-1] < rnd \le ACC\_\gamma^{(v)}[i],
  \]
which again is solved by binary search;
  \item \textbf{Step 3}: Return $N_v[i]$.
\end{itemize}
Its computational complexity is $O(\log d_v)$.

\textbf{$\text{WeightRandomVertex}(N_u\setminus \{v\}, \rho^{(u, v)})$}:
As alluded, we use $N_u[1, \ldots, d_u]$ to store the neighbors of $u$,
and $ACC\_\alpha^{(u)}[1, \ldots, d_u]$ to store $ACC\_\alpha^{(u)}[i] = \sum_{j=1}^i (d_{N_u[j]} - 1)$, $1\le i\le d_u$.
Let $POS_{u, v}$ be the index of $v$ in $N_u[1, \ldots, d_u]$,
i.e., $N_u[POS_{u, v}]=v$.
Then, function $\text{WeightRandomVertex}(N_u\setminus \{v\}, \rho^{(u, v)})$ consists of the following steps:
\begin{itemize}
  \item \textbf{Step 1}: Select number $rnd$ from
  $  \left\{1, \ldots, ACC\_\alpha^{(u)}[d_u]\right\}\setminus \left\{ACC\_\alpha^{(u)}[POS_{u, v}-1] + 1, \ldots, ACC\_\alpha^{(u)}[POS_{u, v}]\right\}$ at random;
  \item \textbf{Step 2}: Find $i$ such that
  \[
  ACC\_\alpha^{(u)}[i-1] < rnd \le ACC\_\alpha^{(u)}[i],
  \]
which is solved by binary search;
  \item \textbf{Step 3}: Return $N_u[i]$.
\end{itemize}
Its computational complexity is $O(\log d_u)$.

\subsection*{Proof of Theorem~\ref{theorem:estimatecardinality}}
For $1\le i\le r$ and $1\le j\le K$, we have
\[
P(X_j\in S_i) = \sum_{s\in S_i} P(X_j=s, s\in S_i)=p_i n_i.
\]
Since $X_1, \ldots, X_K$ are sampled independently,
the random variable $\sum_{j=1}^K \mathbf{1}(X_j\in S_i)$ follows the binomial distribution with parameters $K$ and $p_i n_i$.
Then, the expectation and variance of $\sum_{j=1}^K \mathbf{1}(X_j\in S_i)$ are
\[
\mathbb{E} \left(\sum_{j=1}^K \mathbf{1}(X_j\in S_i)\right) = K p_i n_i,
\]
\[
\text{Var} \left(\sum_{j=1}^K \mathbf{1}(X_j\in S_i)\right) = K p_i n_i (1 - p_i n_i).
\]
Therefore, the expectation and variance of $\hat n_i$ are computed as
\begin{equation*}
\mathbb{E} (\hat n_i) = \mathbb{E} \left(\frac{\sum_{j=1}^K \mathbf{1}(X_j\in S_i)}{K p_i}\right) = n_i,
\end{equation*}
\begin{equation*}
\text{Var} (\hat n_i) = \text{Var}\left(\frac{\sum_{j=1}^K \mathbf{1}(X_j\in S_i)}{K p_i}\right) = \frac{n_i}{K} \left(\frac{1}{p_i} - n_i\right).
\end{equation*}

For $i\ne j$ and $1\le i, j\le r$,
the covariance of $\hat n_i$ and $\hat n_j$ is
\begin{eqnarray}
&&\text{Cov}(\hat n_i, \hat n_j)\nonumber\\
&=& \text{Cov} \left(\frac{\sum_{t=1}^{K} \mathbf{1}(X_t\in S_i)}{K p_i}, \frac{\sum_{l=1}^{K} \mathbf{1}(X_l\in S_j)}{K p_j}\right)\nonumber \\
&=& \frac{\text{Cov} (\sum_{t=1}^{K} \mathbf{1}(X_t\in S_i), \sum_{l=1}^{K} \mathbf{1}(X_l\in S_j))}{K^2 p_i p_j}\nonumber \\
&=& \frac{ \sum_{t=1}^{K} \sum_{l=1}^{K} \text{Cov}(\mathbf{1}(X_t\in S_i), \mathbf{1}(X_l\in S_j))}{K^2 p_i p_j}\nonumber \\
&=& \frac{ \sum_{t=1}^{K} \text{Cov}(\mathbf{1}(X_t\in S_i), \mathbf{1}(X_t\in S_j))}{K^2 p_i p_j}\nonumber \\
&=& -\frac{n_i n_j}{K}.\nonumber
\end{eqnarray}
In the derivation above, we use
\[
\text{Cov}(\mathbf{1}(X_t\in S_i), \mathbf{1}(X_l\in S_j)) = 0, \quad t\ne l,
\]
\begin{equation*}
\begin{split}
&\text{Cov}(\mathbf{1}(X_t\in S_i), \mathbf{1}(X_t\in S_j))\\
=&\mathbb{E}(\mathbf{1}(X_t\in S_i) \mathbf{1}(X_t\in S_j)) - \mathbb{E}(\mathbf{1}(X_t\in S_i)) \mathbb{E}(\mathbf{1}(X_t\in S_j))\\
=&0-p_i n_i p_j n_j\\
=&-p_i p_j  n_i n_j.
\end{split}
\end{equation*}

\begin{figure*}[htb]
\center
\includegraphics[width=\textwidth]{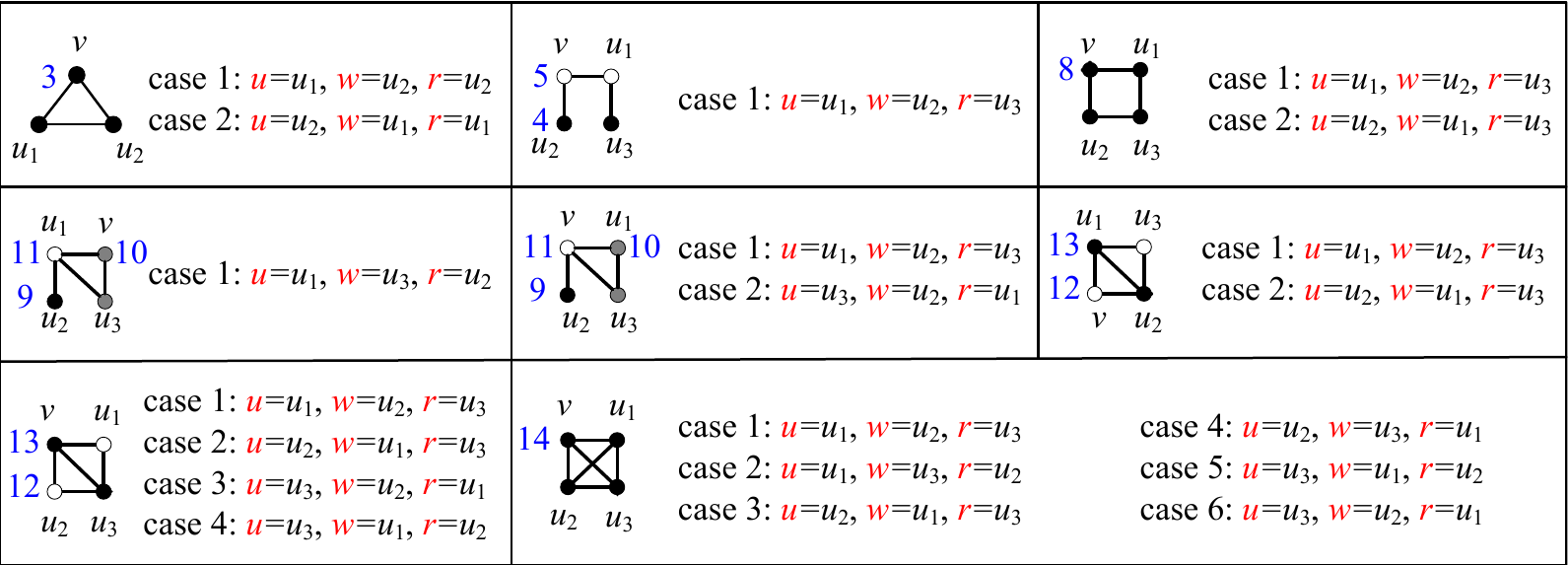}
\caption{The ways of Randgraf-4-1 sampling a CIS that includes $v$ in different orbits. Numbers in blue are orbit IDs. $u$, $w$, and $r$ in red are the variables in Algorithm~\ref{alg:randgraf-4-1}.}
\label{fig:casegraph4-1}
\end{figure*}

\begin{figure*}[htb]
\center
\includegraphics[width=\textwidth]{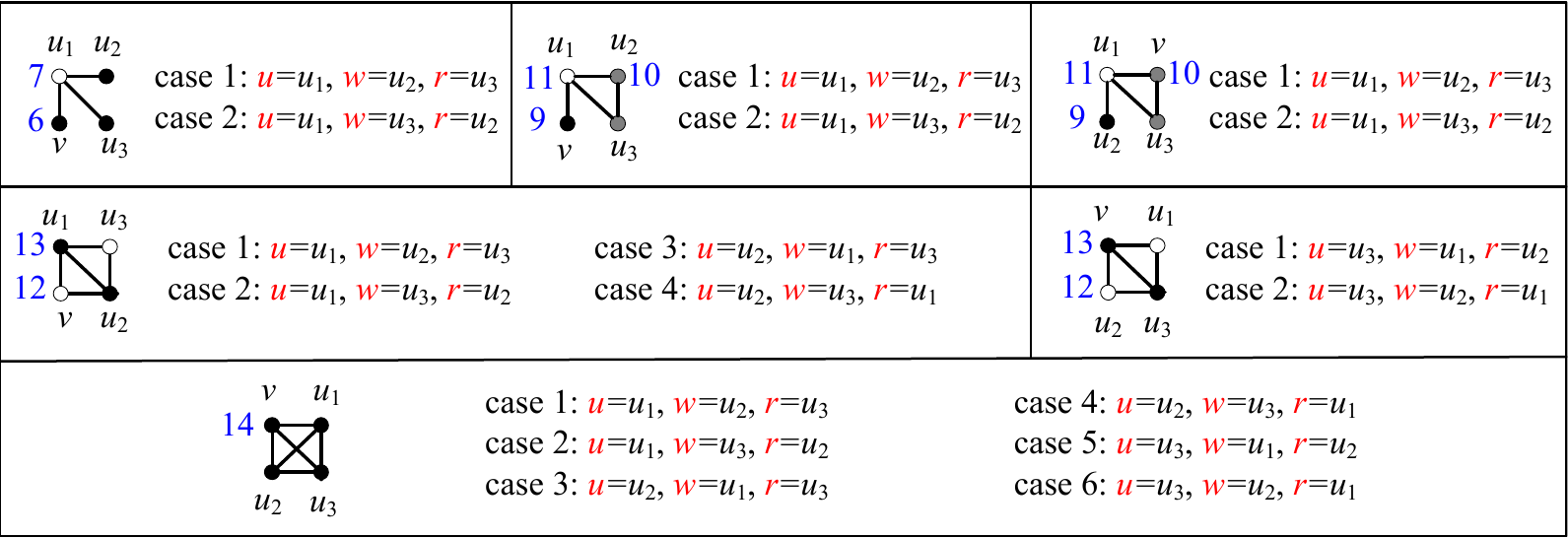}
\caption{The ways of Randgraf-4-2 sampling a CIS that includes $v$ in different orbits. Numbers in blue are orbit IDs. $u$, $w$, and $r$ in red are the variables in Algorithm~\ref{alg:randgraf-4-2}.}
\label{fig:casegraph4-2}
\end{figure*}

\begin{figure*}[htb]
\center
\includegraphics[width=\textwidth]{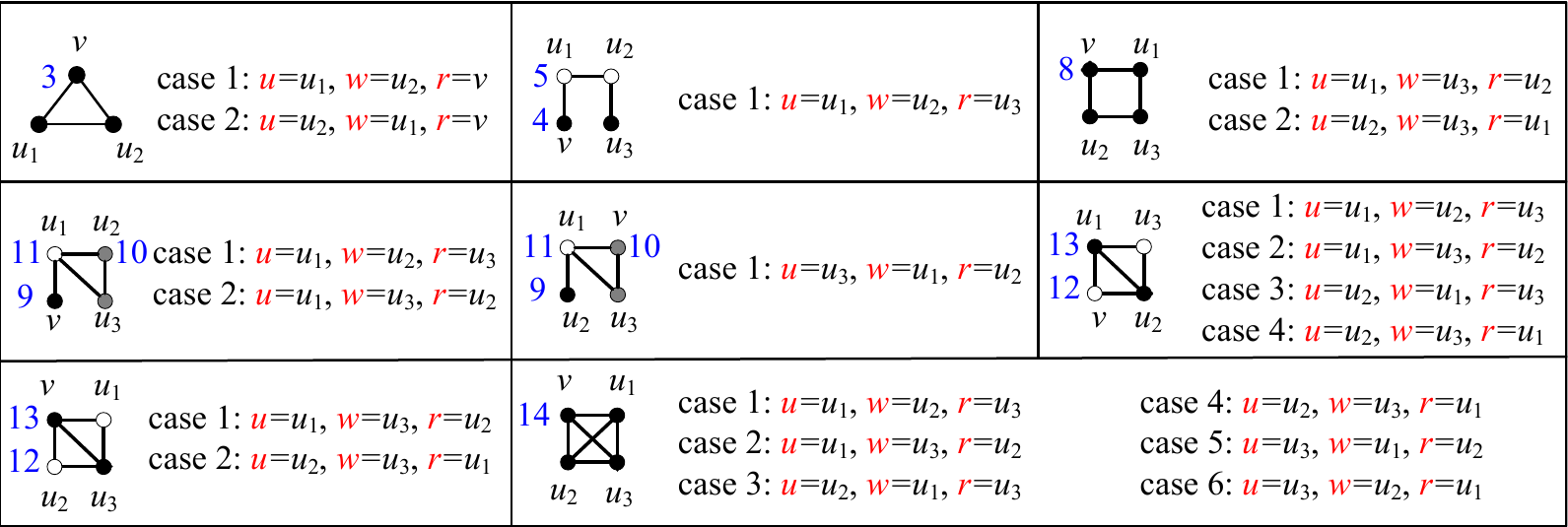}
\caption{The ways of Randgraf-4-3 sampling a CIS that includes $v$ in different orbits. Numbers in blue are orbit IDs. $u$, $w$, and $r$ in red are the variables in Algorithm~\ref{alg:randgraf-4-3}.}
\label{fig:casegraph4-3}
\end{figure*}
\subsection*{Proof of Theorem~\ref{theorem:prob_randgraf-3-1}}
The number of selections of variables $u$ and $w$ in Algorithm~\ref{alg:randgraf-3-1} is ${d_v \choose 2} \times 2! = 2\phi_v$.
For a CIS $s$ consisting three nodes $v$, $u_1$, and $u_2$,
when $s$ includes $v$ in orbit 2 or 3,
Randgraf-3-1 has two ways to sample $s$:
(1) $u = u_1$ and $w = u_2$; (2) $u = u_2$ and $w = u_1$.
Each happens with probability $\frac{1}{d_v} \times \frac{1}{d_v-1} = \frac{1}{2\phi_v}$.
Otherwise, Randgraf-3-1 is not able to sample $s$.
Therefore, we have $p_1^{(3,1)} = 0$, $p_2^{(3,1)}=\frac{1}{\phi_v}$, and $p_3^{(3,1)}=\frac{1}{\phi_v}$.

\subsection*{Proof of Theorem~\ref{theorem:prob_randgraf-3-2}}
The number of selections of variables $u$ and $w$ in Algorithm~\ref{alg:randgraf-3-2} is $\varphi_v = \sum_{u\in N_v} (d_u - 1)$.
For a CIS $s$ including $v$ in orbit 1, Randgraf-3-2 has only one way to sample $s$,
which happens with probability $\alpha_u^{(v)} \times \frac{1}{d_u-1} = \frac{1}{\varphi_v}$.
When $s$ including $v$ in orbit 3, similar to Randgraf-3-1,  Randgraf-3-2 has two different ways to sample $s$,
where each happens with probability $\frac{1}{\varphi_v}$.
When $s$ including $v$ in orbit 2, Randgraf-3-2 is not able to sample it.
Therefore, we have $p_1^{(3,2)} = \frac{1}{\varphi_v}$, $p_2^{(3,2)}=0$, and $p_3^{(3,2)}=\frac{2}{\varphi_v}$.

\subsection*{Proof of Theorem~\ref{theorem:prob_randgraf-4-1}}
The number of selections of variables $u$, $w$, and $r$ in Algorithm~\ref{alg:randgraf-4-1} is $(d_v - 1) \sum_{u\in N_v} (d_u - 1) = \Phi_v^{(1)}$.
As shown in Fig.~\ref{fig:casegraph4-1}, Randgraf-4-1 has
2, 1, 2, 1, 2, 2, 4, and 6 ways to sample a 3- or 4-node CIS $s$
including $v$ in orbits 3, 5, 8, 10, 11, 12, 13, and 14 respectively.
Each way happens with probability $\alpha_u^{(v)} \times \frac{1}{d_u - 1} \times \frac{1}{d_v-1} = \frac{1}{\Phi_v^{(1)}}$.
When $s$ includes $v$ in the other orbits,
Randgraf-4-1 cannot not sample $s$.
Therefore, we have
$p_1^{(4,1)} = p_2^{(4,1)} = p_4^{(4,1)} = p_6^{(4,1)} = p_7^{(4,1)} = p_9^{(4,1)} = 0$,
$p_3^{(4,1)}=\frac{2}{\Phi_v^{(1)}}$,
$p_5^{(4,1)}=\frac{1}{\Phi_v^{(1)}}$,
$p_8^{(4,1)}=\frac{2}{\Phi_v^{(1)}}$,
$p_{10}^{(4,1)}=\frac{1}{\Phi_v^{(1)}}$,
$p_{11}^{(4,1)}=\frac{2}{\Phi_v^{(1)}}$,
$p_{12}^{(4,1)}=\frac{2}{\Phi_v^{(1)}}$,
$p_{13}^{(4,1)}=\frac{4}{\Phi_v^{(1)}}$,
and $p_{14}^{(4,1)}=\frac{6}{\Phi_v^{(1)}}$.

\subsection*{Proof of Theorem~\ref{theorem:prob_randgraf-4-2}}
The number of selections of variables $u$, $w$, and $r$ in Algorithm~\ref{alg:randgraf-4-2} is $\sum_{u\in N_v}(d_u - 1)(d_u - 2) = 2 \Phi_v^{(2)}$.
As shown in Fig.~\ref{fig:casegraph4-2}, Randgraf-4-2 has
2, 2, 2, 4, 2, and 6 ways to sample a 4-node CIS $s$
including $v$ in orbits 6, 9, 10, 12, 13, and 14 respectively.
Each way happens with probability $\beta_u^{(v)} \times \frac{1}{d_u - 1} \times \frac{1}{d_u - 2} = \frac{1}{2\Phi_v^{(2)}}$.
When $s$ includes $v$ in the other orbits,
Randgraf-4-2 is not able to sample $s$.
Therefore, we have $p_1^{(4,2)} = p_2^{(4,2)} = p_3^{(4,2)} = p_4^{(4,2)} = p_5^{(4,2)} = p_7^{(4,2)} = p_8^{(4,2)} = p_{11}^{(4,2)} = 0$,
$p_6^{(4,2)}=\frac{1}{\Phi_v^{(2)}}$,
$p_9^{(4,2)}=\frac{1}{\Phi_v^{(2)}}$,
$p_{10}^{(4,2)}=\frac{1}{\Phi_v^{(2)}}$,
$p_{12}^{(4,2)}=\frac{2}{\Phi_v^{(2)}}$,
$p_{13}^{(4,2)}=\frac{1}{\Phi_v^{(2)}}$,
and $p_{14}^{(4,2)}=\frac{3}{\Phi_v^{(2)}}$.

\subsection*{Proof of Theorem~\ref{theorem:prob_randgraf-4-3}}
The number of selections of variables $u$, $w$, and $r$ in Algorithm~\ref{alg:randgraf-4-3} is
$\sum_{u\in N_v} \sum_{w\in N_u-\{v\}} (d_w - 1) = \sum_{u\in N_v}(\varphi_u - d_v + 1) = \Phi_v^{(3)}$.
As shown in Fig.~\ref{fig:casegraph4-3}, Randgraf-4-3 has
2, 1, 2, 2, 1, 4, 2, and 6 ways to sample a 4-node CIS $s$
including $v$ in orbits 3, 4, 8, 9, 10, 12, 13, and 14 respectively.
Each way happens with probability $\gamma_u^{(v)} \times \rho_w^{(u, v)} \times \frac{1}{d_w - 1} = \frac{1}{\Phi_v^{(3)}}$.
When $s$ includes $v$ in the other orbits,
Randgraf-4-3 is not able to sample $s$.
Therefore, we have $p_1^{(4,3)} = p_2^{(4,3)} = p_5^{(4,3)} = p_6^{(4,3)} = p_7^{(4,3)} = p_{11}^{(4,3)} = 0$,
$p_3^{(4,3)}=\frac{2}{\Phi_v^{(3)}}$,
$p_4^{(4,3)}=\frac{1}{\Phi_v^{(3)}}$,
$p_8^{(4,3)}=\frac{2}{\Phi_v^{(3)}}$,
$p_9^{(4,3)}=\frac{2}{\Phi_v^{(3)}}$,
$p_{10}^{(4,3)}=\frac{1}{\Phi_v^{(3)}}$,
$p_{12}^{(4,3)}=\frac{4}{\Phi_v^{(3)}}$,
$p_{13}^{(4,3)}=\frac{2}{\Phi_v^{(3)}}$,
and $p_{14}^{(4,3)}=\frac{6}{\Phi_v^{(3)}}$.

\subsection*{Proof of Theorem~\ref{theorem:prob_randgraf-4-4}}
The number of selections of variables $u$, $w$, and $r$ in Algorithm~\ref{alg:randgraf-4-4} is ${d_v \choose 3} \times 3! = 6\Phi_v^{(4)}$.
For a CIS $s$ consisting four nodes $v$, $u_1$, $u_2$, and $u_3$,
when $s$ includes $v$ in orbit 7, 11, 13, or 14,
Randgraf-4-4 has six ways to sample $s$:
(1) $u = u_1$, $w = u_2$, $r = u_3$;
(2) $u = u_1$, $w = u_3$, $r = u_2$;
(3) $u = u_2$, $w = u_1$, $r = u_3$;
(4) $u = u_2$, $w = u_3$, $r = u_1$;
(5) $u = u_3$, $w = u_1$, $r = u_2$;
(6) $u = u_3$, $w = u_2$, $r = u_1$.
Each one happens with probability $\frac{1}{d_v} \times \frac{1}{d_v-1} \times \frac{1}{d_v-2} = \frac{1}{6\Phi_v^{(4)}}$.
When $s$ includes $v$ in the other orbits,
Randgraf-4-4 is not able to sample $s$.
Therefore, we have $p_1^{(4,4)} = p_2^{(4,4)} = p_3^{(4,4)} = p_4^{(4,4)} = p_5^{(4,4)} = p_6^{(4,4)} = p_8^{(4,4)} = p_9^{(4,4)} = p_{10}^{(4,4)} = p_{12}^{(4,4)} = 0$, and $p_{7}^{(4,4)}=p_{11}^{(4,4)}=p_{13}^{(4,4)} = p_{14}^{(4,4)}=\frac{1}{\Phi_v^{(4)}}$.

\subsection*{Proof of Theorem~\ref{theorem:equation_undirected}}
We easily find that the total number of selections of $u$ and $w$ in Algorithm Randgraf-3-1 is $2\phi_v$.
From the proof of Theorem~\ref{theorem:prob_randgraf-3-1}, we observe:
(1) Randgraf-3-1 has two ways to sample CISes including $v$ in both orbits 2 and 3;
(2) Randgraf-3-1 is not able to sample the other CISes including $v$.
Therefore, we have
$2 d^{(2)}_v + 2 d^{(3)}_v = 2 \phi_v$.

We find that the total number of selections of $u$, $w$, and $r$ in Algorithm Randgraf-4-3 is $\Phi_v^{(3)}$.
From the proof of Theorem~\ref{theorem:prob_randgraf-4-3}, we observe:
(1) Randgraf-4-3 has 2, 1, 2, 2, 1, 4, 2, and 6 way/ways to sample CISes including $v$ in orbits 3, 4, 8, 9, 10, 12, 13, and 14 respectively;
(2) Randgraf-4-3 is not able to sample the other CISes including $v$.
Therefore, we have
\begin{equation*}
\begin{split}
&2 d^{(3)}_v + d^{(4)}_v  + 2 d^{(8)}_v + 2 d^{(9)}_v + d^{(10)}_v + 4 d^{(12)}_v + 2 d^{(13)}_v\\
&+ 6 d^{(14)}_v = \Phi_v^{(3)}.
\end{split}
\end{equation*}

We find that the total number of selections of $u$, $w$, and $r$ in Algorithm Randgraf-4-4 is $6\Phi_v^{(4)}$.
From the proof of Theorem~\ref{theorem:prob_randgraf-4-4}, we observe that
(1) Randgraf-4-4 has 6 ways to sample CISes including $v$ in orbits 7, 11, 13, and 14 respectively;
(2) Randgraf-4-4 is not able to sample the other CISes including $v$.
Thus, we have
\begin{equation*}
d^{(7)}_v + d^{(11)}_v + d^{(13)}_v + d^{(14)}_v = \Phi_v^{(4)}.
\end{equation*}

\subsection*{Proof of Theorem~\ref{theorem:error_undirected_GDV}}
According to Theorems~\ref{theorem:estimatecardinality} and~\ref{theorem:prob_randgraf-3-2}, we have
\[
\text{Var}(\hat{d}^{(1)}_v) = \frac{d^{(1)}_v}{K^{(3,2)}}\left(\frac{1}{p_1^{(3,2)}} - d^{(1)}_v\right).
\]
According to Theorems~\ref{theorem:estimatecardinality} and~\ref{theorem:prob_randgraf-4-1}, we have
\[
\text{Var}(\hat{d}^{(i)}_v) = \frac{d^{(i)}_v}{K^{(4,1)}}\left(\frac{1}{p_i^{(4,1)}} - d^{(i)}_v\right), i\in \{5, 8, 11\}.
\]
According to Theorems~\ref{theorem:estimatecardinality} and~\ref{theorem:prob_randgraf-4-2}, we have
\[
\text{Var}(\hat{d}^{(i)}_v) = \frac{d^{(i)}_v}{K^{(4,2)}}\left(\frac{1}{p_i^{(4,2)}} - d^{(i)}_v\right), i\in \{6, 9\},
\]
By Theorem~\ref{theorem:mixestimators} and the definition of  $\hat{d}^{(3)}_v$, $\hat{d}^{(10)}_v$, $\hat{d}^{(12)}_v$, $\hat{d}^{(13)}_v$, and $\hat{d}^{(14)}_v$ in Eqs.~(\ref{eq:undirectedhatd3mix}) and~(\ref{eq:undirectedhatdmix}), we have
\begin{equation*}
\begin{split}
\text{Var}(\hat{d}^{(i)}_v) &= \text{Var}\left(\frac{\text{Var}(\tilde{d}^{(i)}_v) \check{d}^{(i)}_v + \text{Var}(\check{d}^{(i)}_v) \tilde{d}^{(i)}_v}{\text{Var}(\check{d}^{(i)}_v) + \text{Var}(\tilde{d}^{(i)}_v)}\right)\\
&= \frac{\text{Var}(\tilde{d}^{(i)}_v) \text{Var}(\check{d}^{(i)}_v)}{\text{Var}(\check{d}^{(i)}_v) + \text{Var}(\tilde{d}^{(i)}_v)}, \qquad i\in \{3, 10, 12, 13, 14\}.
\end{split}
\end{equation*}
In the above derivation, the last equation holds because $\check{d}^{(i)}_v$ and $\tilde{d}^{(i)}_v$ are independent,
which can be easily obtained from their definition in Eqs.~(\ref{eq:varcheck_d_3}),~(\ref{eq:vartilde_d_3}),~(\ref{eq:varcheck_d_10}), and~(\ref{eq:vartilde_d_10}).

For $\text{Var}(\hat{d}^{(2)}_v)$, we have
\[
\text{Var}(\hat{d}^{(2)}_v) = \text{Var}(\phi_v -  \hat d^{(3)}_v) = \text{Var}(\hat{d}^{(3)}_v).
\]
By the definition of $\hat{d}^{(4)}_v$ and $\hat{d}^{(7)}_v$, we easily proof that the formulas of their variances are
\begin{equation*}
\begin{split}
\text{Var}(\hat{d}^{(4)}_v) &= \sum_{j\in \{3, 8, 9, 10, 12, 13, 14\}} \chi_j^2 \text{Var}(\hat{d}^{(j)}_v)\\
&+\sum_{j, k\in \{3, 8, 9, 10, 12, 13, 14\}\wedge j\ne l} \chi_j \chi_l \text{Cov}(\hat{d}^{(j)}_v, \hat{d}^{(l)}_v).
\end{split}
\end{equation*}

\begin{equation*}
\begin{split}
\text{Var}(\hat{d}^{(7)}_v) &= \text{Var}(\hat{d}^{(11)}_v) + \text{Var}(\hat{d}^{(13)}_v) + \text{Var}(\hat{d}^{(14)}_v)\\
&+ \sum_{j, l\in\{11,13,14\}\wedge j\ne l} \text{Cov}(\hat{d}^{(j)}_v, \hat{d}^{(l)}_v),
\end{split}
\end{equation*}

The covariances in the above formulas of $\text{Var}(\hat{d}^{(4)}_v)$ and $\text{Var}(\hat{d}^{(7)}_v)$ are computed as

1. When $j, l\in \{5, 8, 11\}$ and $i\ne l$, by the definition of $\hat{d}^{(j)}_v$ in Eq.~(\ref{eq:hat_d_single}) and Theorem~\ref{theorem:estimatecardinality},
we have $\text{Cov}(\hat{d}^{(j)}_v, \hat{d}^{(l)}_v) = -\frac{d^{(j)}_v d^{(l)}_v}{K^{(4,1)}}$.

2. When $j\in \{5, 8, 11\}$, by the definition of $\hat{d}^{(3)}_v$ and $\hat{d}^{(j)}_v$ in Eqs.~(\ref{eq:undirectedhatd3mix}) and~(\ref{eq:hat_d_single}), we have
\begin{equation*}
\begin{split}
\text{Cov}(\hat{d}^{(3)}_v, \hat{d}^{(j)}_v) &= \text{Cov}(\lambda^{(3,1)}_v \check{d}^{(3)}_v + \lambda^{(3,2)}_v \tilde{d}^{(3)}_v, \hat{d}^{(j)}_v)\\
 &= \lambda^{(3,1)}_v\text{Cov}(\check{d}^{(3)}_v, \hat{d}^{(j)}_v) + \lambda^{(3,2)}_v \text{Cov}(\tilde{d}^{(3)}_v, \hat{d}^{(j)}_v).
\end{split}
\end{equation*}
Since $\tilde{d}^{(3)}_v$ and $\hat{d}^{(j)}_v$ are computed based independent samples generated by Randgraf-3-1 and Randgraf-4-1 respectively,
we have $\text{Cov}(\check{d}^{(3)}_v, \hat{d}^{(j)}_v)=0$.
From Theorem~\ref{theorem:estimatecardinality}, we have $\text{Cov}(\tilde{d}^{(3)}_v, \hat{d}^{(j)}_v)=-\frac{d^{(3)}_v d^{(j)}_v}{K^{(4,1)}}$.
Therefore, we have  $\text{Cov}(\hat{d}^{(3)}_v, \hat{d}^{(j)}_v)=-\frac{\lambda^{(3,1)}_v d^{(3)}_v d^{(j)}_v}{K^{(4,1)}}$.

3. When $j, l\in \{6, 9\}$ and $j\ne l$, by the definition of $\hat{d}^{(j)}_v$ in Eq.~(\ref{eq:hat_d_single}) and Theorem~\ref{theorem:estimatecardinality}, we have $\text{Cov}(\hat{d}^{(j)}_v, \hat{d}^{(l)}_v) = -\frac{d^{(j)}_v d^{(l)}_v}{K^{(4,2)}}$.

4. When $j, l\in \{10, 12, 13, 14\}$ and $j\ne l$, by the definition of $\hat{d}^{(j)}_v$ in Eq.~(\ref{eq:undirectedhatdmix}), we have
$\text{Cov}(\hat{d}^{(j)}_v, \hat{d}^{(l)}_v)=  \text{Cov}(\lambda^{(j,1)}_v \check{d}^{(j)}_v + \lambda^{(j,2)}_v \tilde{d}^{(j)}_v, \lambda^{(l,1)}_v \check{d}^{(l)}_v + \lambda^{(l,2)}_v \tilde{d}^{(l)}_v)$.
By the definitions of $\check{d}^{(j)}_v$ and $\tilde{d}^{(j)}_v$ in Eqs.~(\ref{eq:varcheck_d_10}) and~(\ref{eq:vartilde_d_10}),
we find that $\check{d}^{(j)}_v$ and $\tilde{d}^{(l)}_v$ are independent, and $\tilde{d}^{(j)}_v$ and $\check{d}^{(l)}_v$ are independent.
Moreover, from Theorem~\ref{theorem:estimatecardinality}, we have $\text{Cov}(\check{d}^{(j)}_v,\check{d}^{(l)}_v) = -\frac{d^{(j)}_v d^{(l)}_v}{K^{(4,1)}}$
and $\text{Cov}(\tilde{d}^{(j)}_v,\tilde{d}^{(l)}_v) = -\frac{d^{(j)}_v d^{(l)}_v}{K^{(4,2)}}$.
Therefore, we have
$\text{Cov}(\hat{d}^{(j)}_v, \hat{d}^{(l)}_v)= -\sum_{k=1,2} \frac{\lambda^{(j,k)}_v \lambda^{(l,k)}_v d^{(j)}_v d^{(l)}_v}{K^{(4,k)}}$.

5. When $j\in \{3, 5, 8, 11\}$ and $l\in \{6, 9\}$, $\hat{d}^{(j)}_v$ and $\hat{d}^{(l)}_v$ are independent because $\hat{d}^{(j)}_v$ are computed based on samples generated by Randgraf-3-1 and Randgraf-4-1, while $\hat{d}^{(l)}_v$ are computed based on samples generated by Randgraf-4-2.

6. When $j \in \{5, 8, 11\}$ and $l\in \{10, 12, 13, 14\}$, we have
$\text{Cov}(\hat{d}^{(j)}_v, \hat{d}^{(l)}_v)=  \text{Cov}(\hat d^{(j)}_v , \lambda^{(l,1)}_v \check{d}^{(l)}_v + \lambda^{(l,2)}_v \tilde{d}^{(l)}_v)$ by the definition of $\hat{d}^{(j)}_v$ in Eq.~(\ref{eq:undirectedhatdmix}).
By the definition of $\hat d^{(j)}_v$ and $\tilde{d}^{(l)}_v$ in Eqs.~(\ref{eq:hat_d_single}) and~(\ref{eq:vartilde_d_10}),
we find that $\hat d^{(j)}_v$ and $\tilde{d}^{(l)}_v$ are independent.
Moreover, by Theorem~\ref{theorem:estimatecardinality} and the definition of $\hat d^{(j)}_v$ and $\check{d}^{(l)}_v$ in Eqs.~(\ref{eq:hat_d_single}) and~(\ref{eq:varcheck_d_10}), we have $\text{Cov}(\hat{d}^{(j)}_v,\check{d}^{(l)}_v) = -\frac{d^{(j)}_v d^{(l)}_v}{K^{(4,1)}}$.
Therefore, we have
$\text{Cov}(\hat{d}^{(j)}_v, \hat{d}^{(l)}_v) = -\frac{\lambda^{(l,1)}_v d^{(j)}_v d^{(l)}_v}{K^{(4,1)}}$.

7. When $j \in \{6, 9\}$ and $l\in \{10, 12, 13, 14\}$, by the definition of $\hat{d}^{(j)}_v$ in Eq.~(\ref{eq:undirectedhatdmix}), we have
$\text{Cov}(\hat{d}^{(j)}_v, \hat{d}^{(l)}_v)=  \text{Cov}(\hat d^{(j)}_v , \lambda^{(l,1)}_v \check{d}^{(l)}_v + \lambda^{(l,2)}_v \tilde{d}^{(l)}_v)$.
By the definition of $\hat d^{(j)}_v$ and $\check{d}^{(l)}_v$ in Eqs.~(\ref{eq:hat_d_single}) and~(\ref{eq:varcheck_d_10}),
we find that $\hat d^{(j)}_v$ and $\check{d}^{(l)}_v$ are independent.
Moreover, by Theorem~\ref{theorem:estimatecardinality} and the definition of $\hat d^{(j)}_v$ and $\tilde{d}^{(j)}_v$ in Eqs.~(\ref{eq:hat_d_single}) and~(\ref{eq:varcheck_d_10}), we have $\text{Cov}(\hat{d}^{(j)}_v,\tilde{d}^{(l)}_v) = -\frac{d^{(j)}_v d^{(l)}_v}{K^{(4,2)}}$.
Therefore, we have
$\text{Cov}(\hat{d}^{(j)}_v, \hat{d}^{(l)}_v) = -\frac{\lambda^{(l,2)}_v d^{(j)}_v d^{(l)}_v}{K^{(4,2)}}$.

8. When $j\in \{10, 12, 13, 14\}$, by the definition of $\hat{d}^{(3)}_v$ and $\hat{d}^{(j)}_v$ in Eqs.~(\ref{eq:undirectedhatd3mix}) and~(\ref{eq:undirectedhatdmix}), we have
$\text{Cov}(\hat{d}^{(j)}_v, \hat{d}^{(3)}_v)=  \text{Cov}(\lambda^{(j,1)}_v \check{d}^{(j)}_v + \lambda^{(j,2)}_v \tilde{d}^{(j)}_v, \lambda^{(3,1)}_v \check{d}^{(3)}_v + \lambda^{(3,2)}_v \tilde{d}^{(3)}_v)$.
By the definition of $\check{d}^{(3)}_v$, $\tilde{d}^{(3)}_v$, $\check{d}^{(j)}_v$, and $\tilde{d}^{(j)}_v$ in
Eqs.~(\ref{eq:varcheck_d_3}),~(\ref{eq:vartilde_d_3}),~(\ref{eq:varcheck_d_10}), and~(\ref{eq:vartilde_d_10}),
we find that $\tilde d^{(3)}_v$ and $\tilde{d}^{(j)}_v$ are independent,
$\tilde d^{(3)}_v$ and $\check{d}^{(j)}_v$ are independent,
and $\check d^{(3)}_v$ and $\tilde{d}^{(j)}_v$ are independent.
We also have $\text{Cov}(\check{d}^{(3)}_v,\check{d}^{(j)}_v) = -\frac{d^{(3)}_v d^{(j)}_v}{K^{(4,1)}}$ from Theorem~\ref{theorem:estimatecardinality}.
Thus, we have
$\text{Cov}(\hat{d}^{(j)}_v, \hat{d}^{(3)}_v) = -\frac{\lambda^{(3,1)}_v \lambda^{(j,1)}_v d^{(3)}_v d^{(j)}_v}{K^{(4,1)}}$.

\balance
\bibliographystyle{unsrt}

\end{document}